\documentclass[aps,prd]{revtex4}
\usepackage{amssymb,amsmath,amsfonts,comment,float,latexsym,graphicx,epstopdf}
\usepackage{color}
\usepackage{fancyvrb}
\usepackage{chngcntr}
\usepackage{etoolbox}

\usepackage{bbm,epsfig}

\usepackage{array}
\usepackage{lipsum}
\usepackage{amssymb}
\usepackage{graphicx}
\usepackage{hepunits}

\newcommand{\be}{\begin{equation}}
\newcommand{\ee}{\end{equation}}
\newcommand{\ba}{\begin{eqnarray}}
\newcommand{\ea}{\end{eqnarray}}

\newcommand{\grts}{\raise.3ex\hbox{$>$\kern-.75em\lower1ex\hbox{$\sim$}}}
\newcommand{\lets}{\raise.3ex\hbox{$<$\kern-.75em\lower1ex\hbox{$\sim$}}}

{\catcode`\|=\active\gdef\Braket#1{\left<\mathcode`\|"8000\let|\bravert {#1}\right>}}

\def\bravert{\egroup\,\vrule\,\bgroup}

\begin{document}

\title{A Three Doublet Lepton-Specific Model}

\author{Marco Merchand}\email[]{mamerchandmedi@email.wm.edu}
\author{Marc Sher}\email[]{mtsher@wm.edu}

\affiliation{High Energy Theory Group, College of William and Mary, Williamsburg, Virginia 23187, U.S.A.}
\date{\today}

\begin{abstract}

In the lepton-specific version of  two Higgs doublet models, a discrete symmetry is used to couple one Higgs, $\Phi_2$, to quarks and the other, $\Phi_1$, to leptons.  The symmetry eliminates tree level flavor changing neutral currents (FCNC).  Motivated by strong constraints on such currents in the quark sector from meson-antimeson mixing, and by hints of $h \to \mu\tau$ in the lepton sector, we study a simple three Higgs doublet model in which one doublet couples to quarks and the other two to leptons.   Unlike most other studies of three Higgs doublet models, we impose no flavor symmetry and just use a $Z_2$ symmetry to constrain the Yukawa couplings.    We present the model and discuss the various mixing angles.  Constraining the parameters to be consistent with observations of the Higgs boson at the LHC, we study the properties of the charged Higgs boson(s) in the model, focusing on the case in which the charged Higgs is above the top threshold.   It is found that one can have the branching fraction of the charged Higgs into $\tau\nu_\tau$ comparable to  $t\bar{b}$ decay without needing very large values for the ratios of vevs.   One can also get a large branching fraction for the much more easily observable $\mu\nu_\tau$ decay.
\end{abstract}
\maketitle

\section{Introduction}

In 2012, the discovery of the Higgs boson\cite{Aad:2012tfa,Chatrchyan:2012xdj} was announced.   To date, the properties of the Higgs have not significantly deviated from the expectations of the Standard Model.    Nonetheless, many expect new physics to be found at the TeV scale.   The Standard Model has a metastable vacuum, a hierarchy problem and no dark matter candidate.   In most TeV scale extensions of the Standard Model,  including  supersymmetric models \cite{Martin:1997ns}, composite Higgs models \cite{Agashe:2004rs}, twin Higgs models \cite{Chacko:2005pe}, and left-right models \cite{Mohapatra:1974hk}, there are additional Higgs fields.     Thus, considerations of extensions of the Higgs sector are well motivated.

The most studied extension of the Higgs sector is the Two Higgs Doublet Model (2HDM); see Ref. \cite{Branco:2011iw} for an extensive review.     As soon as one extends the Higgs sector, tree level flavor-changing neutral currents (FCNC) become an issue \cite{Paschos:1976ay,Glashow:1976nt}.    These can be avoided by imposing a $Z_2$ discrete symmetry such that all fermions of a given charge couple to the same multiplet.    In the 2HDM, there are four models generally discussed.   In the type I model, all fermions couple to one Higgs (generally taken to be $\Phi_2$), and in the type II model, the Q=2/3 quarks couple to $\Phi_2$ and the $Q=-1/3$ quarks and leptons couple to $\Phi_1$.   In the other two models, the lepton specific and flipped models, the leptons couple to the other Higgs.   For the lepton specific model, in particular, the quarks all couple to $\Phi_2$ and the leptons all couple to $\Phi_1$.  

There are hundreds of papers on these four models.    However, any discovery of a tree level FCNC would invalidate them.   We are intrigued by the recent hints \cite{Khachatryan:2015kon} of a nonzero branching ratio of the light Higgs into $\mu\tau$ with 2.4$\sigma$ significance, with a branching ratio of $0.84\pm 0.38\%$.     More recent results have not seen a signal, but the errors are large, with Ref. \cite{CMS:2016qvi} giving a branching ratio of $0.76\pm 0.82\%$ and Ref. \cite{Aad:2016blu} giving a branching ratio of $0.53\pm 0.51\%$.         If such a signal is confirmed, then the conventional 2HDMs will be excluded and models with a different structure in the lepton sector will be favored.   In the (more likely) event that it is not confirmed, it is still the case that bounds on scalar mediated FCNC in the quark sector are much stronger (due to meson-antimeson mixing)  than those in the lepton sector, and thus we wish to study extensions of the Higgs sector with FCNC in the lepton sector but not in the quark sector.

There have been some studies of alternative 2HDMs that can address lepton flavor-violation.   For example, Altmannshofer, et al. \cite{Altmannshofer:2015esa}, Ghosh, et al. \cite{Ghosh:2015gpa} and Botella et al.  \cite{Botella:2016krk} consider models in which the Standard Model Higgs couples to the third generation and the other Higgs (which may be composite) to the other two, and they study higher-dimensional operators and the effect on the low energy theory.   

Since leptons must transform differently than quarks in this case, the simplest model is to extend the lepton specific model by adding a third doublet, $\Phi_3$, which behaves under the $Z_2$ symmetry exactly as $\Phi_1$.    There will then be only one doublet coupling to quarks and the other two to leptons.    The three doublet model has two physical pairs of charged Higgs fields, two pseudoscalars and three neutral scalars.    

Three Higgs doublet models (3HDMs) have been studied previously.   They allow for a very rich structure of discrete symmetry groups in the Higgs and flavor sectors.   A detailed analysis of these symmetries has been carried out by Ivanov and collaborators, both for discrete symmetries \cite{Ivanov:2012ry,Ivanov:2014doa} and Abelian symmetries \cite{Ivanov:2011ae} and Keus, King and Moretti \cite{Keus:2013hya} focused on flavor symmetries and analyzed Higgs masses and potentials in several models.    A specific model with an $S_3$ symmetry was considered in Refs. \cite{Das:2014fea, Das:2015sca} and one with an $A_4$ symmetry, that also discusses $h\to\mu\tau$ is in Ref.   \cite{Heeck:2014qea}.     Moretti and Yagyu \cite{Moretti:2015cwa} studied perturbative unitarity in 3HDMs in which some of the doublets are inert.    More phenomenological analyses were carried out by Aranda, et al. \cite{Aranda:2013kq,Aranda:2014jua} in the context of a 3HDM with a $\mathbb{Z}_5$ symmetry.   A model with a global $U(1)$ flavor symmetry by Crivellin, DÕAmbrosio and Heeck \cite{Crivellin:2015lwa} not only can account for $h\to\mu\tau$ but can also resolve anomalies in $B\to K\mu\mu$.     The possibility of a light charged Higgs and constraints from B physics in 3HDMs was considered by Akeroyd, et al. \cite{Akeroyd:2016ssd}.    These authors, and a subsequent article by Yagyu  \cite{Yagyu:2016whx} recently considered 3HDMs with no tree level FCNC and studied Higgs boson couplings; these works include a new ``Model Z" in which the three doublets couple to the up-type quarks, down-type quarks and leptons, respectively \cite{Akeroyd:2016ssd}.   A nice review can be found in Yagyu's recent talk \cite{yagyutalk}.      

Our model differs from these in that we do have tree level FCNC, albeit only in the lepton sector,  the discrete symmetry is a simple $Z_2$ and we will focus on charged Higgs masses above the top quark threshold.   In the lepton specific 2HDM, one can only have $\tau\nu$ charged Higgs decays dominate above the top quark threshold if there is a very large ratio of vacuum expectation values and we wish to see if the requirement that there be such a large ratio still holds in the 3HDM.   It will also be noted that the isospin counterpart of $h\rightarrow\mu\tau$ would be $H^+\rightarrow \mu\nu_\tau$, leading to a substantial and observable enhancement of muonic decays of the charged Higgs.

In the next section, we present the model, scalar mass matrices and mixing angles.  In Section 3, we study the charged Higgs bosons, including bounds on the parameters from LHC observations of the light Higgs and focusing on the $\tau\nu$ decay mode which, we will show, can dominate the decays even above the top threshold and in Section 4 we present our conclusions.

\section{ The Model}

The model consists of three Higgs doublets with weak hypercharge $Y=1/2$.    Since FCNC in the quark sector are very small, and we hope to allow tree level FCNC in the lepton sector, quarks and leptons must transform differently under the $Z_2$ symmetry.    In the conventional 2HDMs, one model does treat quarks and leptons differently.  This is the lepton specific model (sometimes referred to as the Type X model).   In this model, the only fields odd under the $Z_2$ are $\Phi_1$ and the right-handed leptons, $e^i_R$.   This then forces $\Phi_1$ to couple only to leptons and $\Phi_2$ only to quarks.   In this simple extension, we introduce a third doublet, $\Phi_3$, which is odd under the $Z_2$.   Thus $\Phi_2$ only couples to quarks, whereas $\Phi_1$ and $\Phi_3$ only couple to leptons.   It is the fact that two doublets couple to leptons that will lead to FCNC in the lepton sector.

\subsection{  Mass matrices and mixing angles}

The scalar potential consistent with the $Z_2$ symmetry (under which $\Phi_1, \Phi_3$ and $e_R^i$ are odd) can be written
\begin{align}
V =& m_{11}^2 \Phi_1^\dagger \Phi_1+m_{22}^2 \Phi_2^\dagger \Phi_2 + m_{33}^2 \Phi_3^\dagger \Phi_3 - m_{13}^2 (\Phi_1^\dagger \Phi_3+\Phi_3^\dagger \Phi_1)+\frac{1}{2}\lambda_{11}(\Phi_1^\dagger \Phi_1)^2 \nonumber \\
   &+ \frac{1}{2}\lambda_{22}(\Phi_2^\dagger \Phi_2)^2+\frac{1}{2}\lambda_{33}(\Phi_3^\dagger \Phi_3)^2+\lambda_{12}\Phi_1^\dagger \Phi_1\Phi_2^\dagger\Phi_2 + \lambda_{13}\Phi_1^\dagger \Phi_1\Phi_3^\dagger\Phi_3  \nonumber \\
   &+ \lambda_{23}\Phi_2^\dagger \Phi_2\Phi_3^\dagger\Phi_3 + \frac{\beta_{12}}{2}\left[(\Phi_1^\dagger \Phi_2)^2 + (\Phi_2^\dagger \Phi_1)^2\right] + \alpha_{12}\Phi_1^\dagger \Phi_2 \Phi_2^\dagger \Phi_1  \nonumber \\
   & + \frac{\beta_{13}}{2}\left[(\Phi_1^\dagger \Phi_3)^2 + (\Phi_3^\dagger \Phi_1)^2\right] + \alpha_{13}\Phi_1^\dagger \Phi_3 \Phi_3^\dagger \Phi_1  \nonumber \\
   &  + \frac{\beta_{23}}{2}\left[(\Phi_2^\dagger \Phi_3)^2 + (\Phi_3^\dagger \Phi_2)^2\right] + \alpha_{23}\Phi_2^\dagger \Phi_3 \Phi_3^\dagger \Phi_2. \label{potential}
\end{align}
In this expression, we have omitted quartic terms odd in either $\Phi_1$ or $\Phi_3$, such as $\Phi_1^\dagger\Phi_1\Phi_1^\dagger\Phi_3$.   This is entirely done for simplicity, and will avoid  needing to solve cubic equations.    This simplification will only affect expressions which explicitly have the $\lambda_{ij}, \alpha_{ij}$ and $\beta_{ij}$.   The phenomenology only depends on the mixing angles and not on these couplings (which realistically can't be measured for decades).    As a result, nothing more will be learned by adding these extra terms. 

We choose parameters such that each doublet acquires a vacuum expectation value (vev).   Some models choose parameters so that one or two of the doublets get zero vev -- these models are inert models and have a dark matter candidate.  Expanding around the minimum we can write
\begin{align}
   \Phi_a  &= \begin{pmatrix}
           \phi_a^{+}\\
           \frac{v_a+\rho_a+i \eta _a}{\sqrt{2}} \\
           \end{pmatrix}, \quad  a=1,2,3. \label{doublets}
  \end{align}
  
With this potential and our choice of minimum, one can find the mass matrices.   This is done in Appendix A.    The charged Higgs, pseudoscalar and scalar mass matrices are all $3\times 3$.   The charged and pseudoscalar mass matrices have a zero eigenvalue, corresponding to the Goldstone bosons.   

Before diagonalizing the mass matrices, it is instructive to recall the diagonalization in the more well-known 2HDMs \cite{Branco:2011iw}.    The minimum of the potential in that model lies along the ray $(v_1,v_2)$.   The mass matrices for the charged scalars and pseudoscalars have a zero eigenvalue, corresponding to the Goldstone bosons.  The mass matrix for the scalars does not have a zero eigenvalue, and one eigenvalue is the $125$ GeV Higgs boson.     If one rotates the basis by a rotation angle given by $\beta\equiv \tan^{-1}(v_2/v_1)$ (which is equivalent to defining a new ``x-axis" along the ray), then one has the Higgs basis, in which only one field gets a vev.  After this rotation, the zero eigenvalues separate out and decouple.    Note that there is only a single rotation angle for both the charged scalar and pseudoscalar matrices, which is not surprising.    Note also that in the 2HDM, the charged scalar and pseudoscalar matrices commute, so they are simultaneously diagonalizable.

In the 3HDM, as can be seen from the mass matrices, the charged scalar and pseudoscalar matrices do not commute, and thus they will not be simultaneously diagonalizable.     Thus, in general, we will have nine different angles (three for each matrix) that diagonalize the $\phi^+$, $\eta$ (pseudoscalar) and the $\rho$ (scalar) matrices.
It turns out that the charged and the pseudoscalar mass matrices have two common angles which diagonalize the matrices, leaving us with a total of seven different angles.   The fact that two of the angles are common is expected.   Two angles are needed to take the general vector $(v_1,v_2,v_3)$ to $(v,0,0)$, and this will in both cases lead to separating out the Goldstone mode.    However, there is still another angle corresponding to an additional rotation about the $(v,0,0)$ axis, and there is no reason that this angle should be the same for the charged and pseudoscalar cases.   

The first rotation matrix that partially diagonalizes the $\phi^+$ and the  $\eta$ matrices  is  the rotation matrix that takes the ray $(v_1,v_2,v_3)$ to $(v,0,0)$.   The first transformation is given by (where $v_{ij}^2 \equiv v_i^2 + v_j^2$)
\begin{equation}
R= \left(\begin{array}{ccc}
\frac{v_1}{v} & \frac{v_2}{v} & \frac{v_3}{v}  \\
-\frac{v_2}{v_{12}} & \frac{v_1}{v_{12}} & 0\\
-\frac{v_1v_3}{v v_{12}} & -\frac{v_2 v_3}{v v_{12}}& \frac{v_{12}}{v}  
\end{array} \right).
\end{equation}  
This is equivalent to rotating the vector $(v_1,v_2,v_3)$ clockwise around the z-axis  by angle $\psi$ and then rotate counter-clockwise around the y-axis by angle $\pi/2-\theta$. The azimuthal and the polar angles are given in terms of the vev's by 
\begin{equation}
\tan{\theta}=\frac{v_{12}}{v_{3}}, \quad \quad \tan{\psi}= \frac{v_2}{v_1}. \label{angles}
\end{equation}
The third rotation, which is the angle of rotation around the x-axis will be called $\beta_1$ and $\beta_2$ for the $\phi^+$ and the $\eta$ matrices respectively. 

For the scalar mass-squared matrix we denote the Euler angles (in the  above convention) that diagonalize it by $\beta_3$, $\theta_3$ and $\psi_3$, where we have replaced the standard notation of performing a rotation by angle $\phi$ first, around the z-axis, by $\beta_3$ instead, in order to avoid confusion with the fields. 
 
 In summary for the charged and the pseudoscalar matrices we perform the following transformation on the fields
 \begin{equation}
 T_{i} = \left( \begin{array}{ccc}
 \cos{\psi}& -\sin{\psi}& 0 \\
 \sin{\psi} & \cos{\psi} & 0\\
0 & 0 & 1
 \end{array} \right)
 \left(\begin{array}{ccc}
 \cos{(\pi/2-\theta)} & 0& -\sin{(\pi/2-\theta)} \\
 0 & 1 & 0\\
\sin{(\pi/2-\theta)} & 0 & \cos{(\pi/2-\theta)}
 \end{array} \right) 
  \left( \begin{array}{ccc}
 1 & 0 & 0 \\
0 & \cos{\beta_{i}} &-\sin{\beta_{i}}\\
0 &\sin{\beta_{i}}& \cos{\beta_{i}} 
 \end{array}\right)  , 
 \end{equation}
where $i=1,2$ and the angles $\theta$ and $\psi$ are given in (\ref{angles}).

For the scalar we write the transformation as the product of three Euler rotations, namely  
\begin{equation}
T_3 = \left( \begin{array}{ccc}
\cos{\psi_3} &-\sin{\psi_3} & 0 \\
\sin{\psi_3}& \cos{\psi_3} & 0\\
0 & 0 & 1
\end{array} \right)
\left( \begin{array}{ccc}
\cos{\theta_3} & 0 & -\sin{\theta_3} \\
0 & 1 & 0\\
\sin{\theta_3} & 0 & \cos{\theta_3}
\end{array}\right)
\left( \begin{array}{ccc}
1 & 0 & 0\\
0 &\cos{\beta_3} & -\sin{\beta_3}\\
0 & \sin{\beta_3} & \cos{\beta_3}
\end{array} \right).
\end{equation}
The field redefinitions are then given by 
\begin{equation}
\begin{pmatrix}
\phi_1^{+}\\
\phi_2^{+}\\
\phi_3^{+}
\end{pmatrix}  = T_1 \begin{pmatrix}
G^+ \\
H_1^+ \\
H_2^+
\end{pmatrix}, \quad \quad 
\begin{pmatrix}
\eta_1\\
\eta_2\\
\eta_3
\end{pmatrix}  = T_2 \begin{pmatrix}
G^0 \\
A_1\\
A_2
\end{pmatrix},  \quad \quad \begin{pmatrix}
\rho_1\\
\rho_2\\
\rho_3
\end{pmatrix}  = T_3 \begin{pmatrix}
h_1 \\
h_2 \\
h_3
\end{pmatrix}. \label{rotations}
\end{equation}
The spectrum of the theory consists of two charged scalars, two pseudo scalars and three neutral scalars.

\subsection{ Gauge boson couplings}

The kinetic term of the Higgs doublets has the form
\begin{equation}
(D_\mu \Phi_i)^\dagger D_\mu \Phi_i,  
\end{equation}
where 
\begin{equation}
D_\mu=\partial_\mu-igW_\mu^a\tau^a-i\frac{g^\prime}{2}B_\mu
\end{equation}
is the covariant derivative and $W_\mu^a$ and $B_\mu$ are the $SU(2)_L$ and $U(1)_Y$ gauge bosons of the electroweak sector respectively. Expanding out in terms of the mass terms given by eq (\ref{rotations}) the Lagrangian contains
\begin{align}
\mathcal{L} \supseteq & -\frac{i g}{2} \sec{\theta_w}\left(\sin{2\theta_w}A^\mu + \cos{2\theta_w}Z^\mu \right) \left(\partial_\mu H_{a-1}^+ H_{a-1}^- -H_{a-1}^+\partial_\mu H_{a-1}^- \right) \\
                      & -\frac{i g}{2}(T_1^T \cdot T_3)_{ab}\left( W_\mu^- \left( H_{a-1}^+ \partial_\mu h_b -\partial_\mu H_{a-1}^+h_b \right) + W_\mu^+ \left(\partial_\mu H_{a-1}^- h_b -H_{a-1}^- \partial_\mu h_b \right) \right)\\
                      & +\frac{g^2}{4} \left(2W_\mu^+W_\mu^- + \sec{\theta_w}^2 \left(Z_\mu \cos{2\theta_w}+A_\mu \sin{2\theta_w} \right)^2\right) H_{a-1}^-H_{a-1}^+ \\
                      & + \frac{g^2}{4}\left(2W_\mu^-W_\mu^+ + Z_\mu^2\sec{\theta_w}^2 \right)(T_3)_{ab}v_a h_b
      \end{align}
where sum over $a,b=1,2,3$ is implied and $H_0^+=G^+$ is the Goldstone boson. From this expression we can read off the couplings of the Higgs particles with the gauge bosons. A list of the relevant couplings to gauge bosons are given at the end of the section.

\subsection{ The quark sector}

As in the lepton-specific model, the RH quarks will couple to $\Phi_2$ and the RH leptons to $\Phi_1$ and $\Phi_3$.
Thus the Yukawa terms in the Lagrangian are
\begin{equation}
-\mathcal{L}_{Yuk} = Y^u_{ij}\bar{Q}^i \tilde{\Phi}_2 u_R^j + Y^d_{ij}\bar{Q}^i\Phi_2d_R^j + \eta^1_{ij}\bar{L}^i\Phi_1e_R^j + \eta^2_{ij}\bar{L}^i\Phi_3e_R^j + h.c. \label{Yuk}
\end{equation}
where
\begin{equation}
\tilde{\Phi}_2 \equiv i\sigma_2 \Phi_2^*,
\end{equation}
and
\begin{equation}
Q^i = \left( \begin{array}{ccc}
u_L^i \\
d_L^i
\end{array} \right),  \quad L_i = \left(\begin{array}{ccc}
\nu_L^i \\
e_L^i 
\end{array}\right).
\end{equation}
Here $i=1,2,3$ are the generation indices.

Quarks only couple to the $\Phi_2$ doublet. The Lagrangian of the quark sector is given by
\begin{equation}
\mathcal{L}_{quark} = \bar{Q}\tilde{\Phi}_2Y^u u_R + \bar{Q}\Phi_2 Y^d d_R.
\end{equation}
Using equation (\ref{doublets}) and expanding out we can write this as 
\begin{align}
\mathcal{L}_{quark} = &\frac{\rho_2}{\sqrt{2}}\left( \bar{u}_L Y^u u_R + \bar{d}_L Y^d d_R \right) - \frac{i \eta_2}{\sqrt{2}}\left(\bar{u}_LY^uu_R - \bar{d}_L Y^d d_R  \right)  \\
& -\phi_2^- \bar{d}_L Y^u u_R + \phi_2^+ \bar{u}_L Y^d d_R + h.c.
\end{align}

This is the same as the conventional lepton-specific model, and one can thus immediately write, as in that model,
\begin{align}
\mathcal{L}_{quark} =& \frac{\rho_2}{v_2}(m^u \bar{u}u+m^d \bar{d}d) - \frac{\eta_2}{v_2}(m^u \bar{u}i\gamma_5 u -m^d \bar{d}i\gamma_5d) \nonumber \\
                      & + \left(- \frac{\sqrt{2}}{v_2}\phi^-_2 \bar{d}_LV_{CKM}^\dagger m^u u_R + \frac{\sqrt{2}}{v_2}\phi^+_2 \bar{u}_L V_{CKM} m^d d_R + h.c. \right),
\end{align}
 $m^u$ and $m^d$ are the entries of the diagonal matrices $\frac{v_2}{\sqrt{2}}M_{u}$ and  $\frac{v_2}{\sqrt{2}}M_{d}$ respectively.

Expressing this in terms of the physical fields is somewhat more complicated, since there are two charged Higgs, two pseudoscalars and three scalars.
In general, one can write
\begin{align}
\mathcal{L}_{quark}= & - \sum_{i=1}^{3} \sum_{f=u,d}\frac{m^f}{v}\left( \xi_{h_i}^f \bar{f}fh_i - i \xi_{A_{i-1}}^f \bar{f}\gamma_5 f A_{i-1} \right) \nonumber\\
                     & + \left\lbrace \sum_{i=1}^{3}\frac{\sqrt{2}V_{ud}}{v}\bar{u}\left( \xi^u_{H^+_{i-1}}m^uP_L + m^d \xi_{H^+_{i-1}}^d P_R \right)dH_{i-1}^+ + h.c. \right\rbrace \label{Lquark}
\end{align}
with $A_0=G^0$, $H_0^+=G^+$ being the Goldstone bosons.  The couplings are given in terms of the matrix elements of the rotations given by equation (\ref{rotations}).

\subsection{ The Higgs basis and the lepton sector}

The lepton sector of the lepton specific 3HDM has the following terms, written in matrix form (with generation indices understood)
\begin{equation}
\mathcal{L}_{Yukawa} \supseteq - \bar{L}\left( \Phi_1 \eta^1 + \Phi_3 \eta^2 \right) e_R + h.c. \label{leptonsector}
\end{equation}
where $\eta^1$ and $\eta^2$ are general complex Yukawa matrices.

We would like to go to the Higgs basis where we make a rotation $(\Phi_1,\Phi_3)\rightarrow (H_1,H_3)$ such that $H_1$ has zero vev and $\langle H_3 \rangle = v_{13}/\sqrt{2}$. This can be accomplished by performing the field redefinition
\begin{equation}
\left(\begin{array}{ccc}
\Phi_1 \\
\Phi_3
\end{array} \right) =\frac{1}{v_{13}} \left(\begin{array}{ccc}
v_3 & v_1 \\
-v_1 & v_3
\end{array} \right) \left(\begin{array}{ccc}
H_1 \\
H_3
\end{array} \right). \label{Hbasis}
\end{equation} 

Following the same notation as in \cite{Branco:2011iw}, we define the matrices
\begin{equation}
N = \frac{1}{\sqrt{2}} \left( v_3 \eta^1 - v_1 \eta^2 \right),
\end{equation}
\begin{equation}
M=\frac{1}{\sqrt{2}} \left(v_1 \eta^1 + v_3 \eta^2 \right).
\end{equation}
So the Lagrangian of the lepton sector, in the Higgs basis, reads
\begin{equation}
\mathcal{L}_{lepton} = -  \frac{\sqrt{2}}{v_{13}} \bar{L}\left( H_1 N + H_3 M \right) e_R. + h.c.
\end{equation}
In the Higgs basis, only the Yukawa couplings of the doublet $H_3$ generate fermion masses, and may be bi-diagonalized so they do not lead to tree-level FCNC.

When passing to the mass basis of the leptons in which the mass matrices are diagonal we need to simultaneously rotate the left-handed and right-handed leptons: 
\begin{equation}
e_R \rightarrow U_R e_R, \quad L \rightarrow U_L L.
\end{equation}
The Lagrangian transforms as
\begin{equation}
-\mathcal{L}_{lepton} = \frac{\sqrt{2}}{v_{13}}\bar{L} \left( U_L^\dagger N U_R H_1 + U_L^\dagger M U_R H_3 \right) e_R. + h.c.
\end{equation}
Naming 
\begin{equation}
U_L^\dagger N U_R=N_d,
\end{equation}
\begin{equation}
U_L^\dagger M U_R = M_d
\end{equation}
where $M_d= diag (m_e,m_\mu,m_\tau)$ is diagonal with real and positive diagonal elements. 

If after bi-diagonalization, the matrix $N_d$ is not diagonal, then there are scalar tree-level flavour changing neutral interactions in the lepton sector, and the lepton FCNC coupling for those interactions are obtained from the entries of $N_d$.

Since $U_R$ is completely unkown and $N$ is arbitrary, the $N_d$ coefficients are arbitrary; in order to look at specific processes, some assumptions must be made about their magnitudes.

Motivated by stringent constraints on flavor-changing couplings involving the first two generations, Cheng and Sher \cite{Cheng:1987rs} showed that if one requires no fine-tuning in the Yukawa matrices, then the
flavor-changing couplings should be of the order of the geometric mean of the Yukawa couplings of the two fermions. In other words, the so-called Cheng-Sher ansatz  is
\begin{equation}
(N_d)_{ij} = k_{ij}\sqrt{m_i m_j}
\end{equation}
where $k_{ij}$ are of order one. Since the most severe bounds on FCNC arise from the first two generations and the Yukawa couplings of the first two generations are small, this ansatz can explain these bounds without requiring huge scalar masses.

Using this ansatz we can write the flavor changing matrix to be approximately of the form
\begin{equation}
N_d =\left( \begin{array}{ccc}
k_{11}m_e & 0 & 0 \\
0 & k_{22}m_\mu & k_{23} \sqrt{m_\mu m_\tau} \\
0 & k_{32} \sqrt{m_\mu m_\tau} & k_{33} m_\tau
\end{array} \right).
\end{equation}
Note that we have not included the $(12), (13),  (21)$ and $(31)$ elements, since FCNC involving electrons are tightly constrained and the coefficients must be small, as shown in Ref. \cite{Diaz:2002uk}.    Using the Cheng-Sher ansatz, the current bound from $\mu\to e\gamma$ requires \cite{Diaz:2002uk} that  $k_{12},k_{13},k_{21},k_{31}$  cannot be much bigger than one. Including these coefficients would make no difference in the physics discussed in the rest of the paper, and thus for simplicity we do not include them here.

Note that we have assumed that the $(12), (13), (21)$ and $(31)$ elements are negligible, due to the fact that a nonzero value for them could lead to too large a value for $\mu\to e\gamma$.   This is consistent with the Cheng-Sher ansatz since these couplings will depend on the electron mass, which is extremely small.

The Lagrangian of the lepton sector in the mass basis reads
\begin{equation}
-\mathcal{L}_{lepton}= \frac{\sqrt{2}}{v_{13}}\bar{L}\left(H_1N_d + H_3 M_d \right)e_R + h.c.. \label{llepton}
\end{equation}

We can separate out the lepton sector Lagrangian into two components. One that includes the FCNC couplings and the other is flavor diagonal
\begin{equation}
\mathcal{L}_{lepton} =\mathcal{L}_{FCNC} +  \mathcal{L}_{diag},
\end{equation}
with
\begin{align}
\mathcal{L}_{diag} =- \frac{\sqrt{2}m_i}{v_{13}} \left( \left( H_1k_{ii} + H_3 \right) \bar{\nu}_L^i e_R^i + \left( H_1k_{ii} + H_3 \right) \bar{e}_L^i e_R^i \right) + h.c. \label{diag}
\end{align}
where the charged component for the first term and the neutral component for the second term are implied for each Higgs doublet.

Expanding out equation (\ref{diag}) in terms of the mass eigenstates we can write in analogous way as eq (\ref{Lquark}) 
 \begin{align}
 \mathcal{L}_{diag} = -  \sum_{f=e,\mu,\tau} \sum_{b=2}^{3} &\frac{\sqrt{2}}{v}m_f  \left\lbrace \left(Z_{H_{b-1}^+}\bar{\nu}_L f_R H_{b-1}^+ + h.c. \right) +  \left( Z_{h_b} \bar{f}f h_b  - Z_{A_{b-1}} i \bar{f} \gamma_5 f A_{b-1}         \right)   \right\rbrace. \label{diagonal} 
\end{align}
and the coupling constant $Z$ can be found on the next section.

Similarly for the FCNC component 
\begin{equation}
\mathcal{L}_{FCNC}= -  \frac{\sqrt{2}k_{23}}{v_{13}}\sqrt{m_\mu m_\tau} \left( H_1 \left( \bar{\nu}_{\tau_L}\mu_R + \bar{\nu}_{\mu_L}  \tau_R \right) + H_1 \left( \bar{\tau}_L \mu_R + \bar{\mu}_L \tau_R \right) \right) + h.c. \label{LFCNC}
\end{equation}

For now we are only interested in couplings which lead to $h\rightarrow \mu \tau$, thus the relevant terms are
\begin{equation}
\mathcal{L}_{FCNC} \supseteq  - \frac{\sqrt{2}}{v_{13}}H_1 \sqrt{m_\mu m_\tau} \left( k_{23} \bar{\mu}_L \tau_R + k_{32}\bar{\tau}_L \mu_R \right) + h.c.
\end{equation}
where the coupling to the neutral component of the doublet is understood.  Expanding in terms of the scalars $h_i$ with $i=1,2,3$, (note  Eq. \ref{rotations}), we can write this as
\begin{equation}
\mathcal{L}_{FCNC} = -  \frac{\sqrt{m_\mu m_\tau}}{v_{13}^2}\left\lbrace v_3 (T_3)_{1i}-v_1 (T_3)_{3i} \right\rbrace h_i \left\lbrace k_{23} \bar{\mu}_L \tau_R + k_{32}\bar{\tau}_L \mu_R \right\rbrace + h.c., \quad i=1,2,3.
\end{equation}
Choosing $k_{ij}=k_{ji}$ and real gives
\begin{equation}
\mathcal{L}_{FCNC} \supseteq  \frac{\sqrt{m_\mu m_\tau}}{v_{13}^2}\left\lbrace v_3 (T_3)_{1i}-v_1 (T_3)_{3i} \right\rbrace k_{23} h_i \left\lbrace \bar{\tau}\mu + \bar{\mu}\tau \right\rbrace  \quad i=1,2,3.
\end{equation}
which gives the relevant couplings.   The flavor changing couplings of the neutral pseudoscalars are found in a similar way, the resulting expression being given by 
\begin{equation}
\mathcal{L}_{FCNC} \supseteq  \frac{i\sqrt{m_\mu m_\tau}}{v_{13}^2}\left\lbrace v_3 (T_2)_{1i}-v_1 (T_2)_{3i} \right\rbrace k_{23} A_{i-1} \left\lbrace \bar{\tau}\gamma_5\mu + \bar{\mu}\gamma_5 \tau \right\rbrace  \quad i=1,2,3.
\end{equation}

It should be noted that the Cheng-Sher ansatz is completely consistent with the hints from CMS.    In the CMS paper\cite{Khachatryan:2015kon}, they plot their results as a function of the Yukawa couplings $Y_{\mu\tau}$ and $Y_{\tau\mu}$, and also include the line in which the product of the two is $m_\mu m_\tau / v^2$.   From this, one can see that the original Cheng-Sher ansatz gives a result slightly below the `observed" value for $k_{\mu\tau}=1$.    Thus, generally the model would predict a branching ratio of the order of a few tenths of a percent, which is consistent with current observations.

One can also discuss the decoupling limit of the model.    In the conventional 2HDM, as discussed in detail by Gunion and Haber \cite{Gunion:2002zf}, one notes that $\cos(\alpha-\beta)=0$ implies that the light Higgs couples to gauge bosons and fermions with the same couplings as in the Standard Model.   This is the decoupling limit, and only the light Higgs gets a vev.    The other Higgs bosons then have vanishing coupling to gauge bosons, but may still have couplings to fermions.  In the decoupling limit, the other Higgs bosons are taken to be sufficiently heavy that they do not affect quark and lepton phenomenology.     In this model, one can see that the limit $$\cos\theta = \sin\theta_3=\cos\psi_3=\cos\psi= 0$$ gives the coupling of the light Higgs to WW and to fermions equal to their Standard Model values and gives vanishing couplings of the heavy Higgs to  $WW$ and $hW$.   This is certainly sufficient to give the decoupling limit, and means increasing precise bounds on deviations of the light Higgs couplings from the Standard Model values will not eliminate the model, but just restrict the parameter-space somewhat.     

The above constraints are certainly sufficient for decoupling, but are not fully necessary.    In order for the light Higgs to have SM-equivalent couplings to gauge bosons  and for the heavy Higgs to have no couplings to gauge bosons, one needs to have $\psi_3 = \psi$ and $\cos \theta=\sin \theta_3$.    This corresponds to the first and third equalities in the above equation.    In order for the light Higgs to have SM-equivalent couplings to the fermions, one must add the condition $\cot\theta=\cos\psi$, which corresponds to $v_1=v_3$.  The above equation is consistent with this, of course.     Note that the constraints in the above paragraph would lead to quark-phobic charged Higgs bosons, but the more general constraint here does not.   Thus, tighter bounds on deviations of the light Higgs couplings from their Standard Model values will not force the charged Higgs to decay primarily leptonically.

\subsection{  Summary of the couplings}
The gauge, quark and lepton couplings are summarized in the Tables below.

\begin{center}
\begin{tabular}{ |c|c|c| } 
 \hline
  & Gauge boson couplings \\
\hline
 $g_{h_1WW}$ & $ \frac{g^2}{2}v \left(\cos{\theta} \sin{\theta_3}+\sin{\theta}\cos{\theta_3}\cos({\psi_3-\psi}) \right)$ \\
 \hline
 $g_{h_2WW}$& $\frac{g^2}{2} v \cos{\theta} \left[ \cos{\beta_3}\sin{(\psi-\psi_3)} +  \sin{\beta_3}(\cot{\theta}\cos{\theta_3}-\sin{\theta_3}\cos{(\psi_3-\psi})) \right]$ \\
 \hline
 $g_{h_3WW}$ & $\frac{g^2}{2}v \cos{\theta} \left[ \sin{\beta_3}\sin{(\psi_3-\psi)}+\cos{\beta_3}\left(\cot{\theta}\cos{\theta_3}-\sin{\theta_3}\cos{(\psi_3-\psi)} \right) \right]$  \\
 \hline
$g_{H_1 h W}$ & $ \frac{ig}{2}\left( \cos{\theta}\cos{\theta_3}\cos{(\psi-\psi_3)}\sin{\beta_1}-\sin{\beta_1}\sin{\theta}\sin{\theta_3}+\cos{\beta_1}\cos{\theta_3}\sin{(\psi-\psi_3)}\right)$  \\
\hline
$g_{H_2 h W}$ & $ \frac{ig}{2}\left( \cos{\beta_1}\left( \cos{\theta}\cos{\theta_3}\cos{(\psi-\psi_3)}-\sin{\theta}\sin{\theta_3} \right)-\cos{\theta_3}\sin{\beta_1}\sin{(\psi-\psi_3)}  \right)$  \\
\hline
\end{tabular}
\end{center}
Table  $1$: A list of Higgs couplings to gauge bosons. The last two expressions are multiplied by $(p-p^\prime)_\mu$.

\begin{center}
\begin{tabular}{ |c|c|c| } 
 \hline
$\times sin{\theta}\sin{\psi}$ & Quark Couplings  \\
\hline
 $\xi_{h_1}^{u,d}$ & $\cos{\theta_3}\sin{\psi_3}$  \\ 
  $\xi_{h_2}^f$ & $\cos{\beta_3}\cos{\psi_3}-\sin{\beta_3}\sin{\theta_3}\sin{\psi_3}$  \\ 
 $\xi_{h_3}^f$ & $-\cos{\psi_3}\sin{\beta_3}-\cos{\beta_3}\sin{\theta_3}\sin{\psi_3}$  \\ 
 $\xi_{A_{1}}^u$ & $-\cos{\psi}\cos{\beta_2}+\cos{\theta}\sin{\psi}\sin{\beta_2}$  \\ 
  $\xi_{A_{2}}^u$ & $\cos{\psi}\sin{\beta_2}+ \sin{\psi}\cos{\theta}\cos{\beta_2}$  \\ 
$\xi_{A_{1}}^d $ & $-\xi_{A_{1}}^u$  \\ 
$\xi_{A_{2}}^d$    &  $-\xi_{A_{2}}^u$   \\
$\xi_{H^+_1}^u$  & $\cos{\psi}\cos{\beta_1}-\cos{\theta}\sin{\psi}\sin{\beta_1}$   \\
$\xi_{H^+_2}^u$  &  $-\cos{\psi}\sin{\beta_1}- \sin{\psi}\cos{\theta}\cos{\beta_1}$\\
$\xi_{H^+_1}^d$  & $-\xi_{H^+_1}^u$ \\
$\xi_{H^+_2}^d$  & $-\xi_{H^+_2}^u$ \\
 \hline
\end{tabular}
\end{center}
Table  $2$: A list of the couplings appearing in equation (\ref{Lquark}). We do not list the couplings of the Goldstone bosons $G^{\pm}$ and $G^0$.

\begin{center}
\begin{tabular}{ |c|c|c| } 
\hline
$\times\frac{v}{k_{23}\sqrt{m_\mu m_\tau}}(1-\sin^2{\theta}\sin^2{\psi})$& FCNC couplings\\
\hline
$h_1 \bar{\mu}\tau$ & $\cos{\theta}\cos{\theta_3}\cos{\psi_3}-\sin{\theta}\cos{\psi}\sin{\theta_3}$\\ $h_2 \bar{\mu}\tau$& $- \left( \sin{\theta}\cos{\psi}\cos{\theta_3}\sin{\beta_3}+\cos{\theta}(\sin{\beta_3}\cos{\psi_3}\sin{\theta_3}+\cos{\beta_3}\sin{\psi_3}) \right)$\\
$h_3 \bar{\mu}\tau$& $-\left[\sin{\theta}\cos{\psi}\cos{\beta_3}\cos{\theta_3} + \cos{\theta}(\cos{\beta_3}\cos{\psi_3}\sin{\theta_3}-\sin{\beta_3}\sin{\psi_3})\right]$ \\
$i A_1 \bar{\mu}\gamma_5\tau$ & $-\left( \cos{\psi}\cos{\beta_2}-\cos{\theta}\sin{\psi}\sin{\beta_2} \right)$\\
$i A_2 \bar{\mu}\gamma_5\tau$ & $-\left(\cos{\psi}\cos{\beta_2} + \cos{\theta}\sin{\psi}\cos{\beta_2} \right)$\\
$\sqrt{2} H_1^+ (\bar{\nu}_\tau P_R \mu + \bar{\nu}_\mu P_R \tau)$            & $-\left( \cos{\psi}\sin{\beta_1} + \cos{\beta_1}\cos{\theta}\sin{\psi} \right)$ \\
$\sqrt{2} H_2^+ (\bar{\nu}_\tau P_R \mu + \bar{\nu}_\mu P_R \tau)$ & $-\cos{\beta_1}\cos{\psi} + \cos{\theta}\sin{\beta_1}\sin{\psi} $ \\
\hline
\end{tabular}
\end{center}
 Table $3$: These are the lepton flavor violating couplings.

 \begin{center}
\scalebox{0.96}{
\begin{tabular}{ |c|c|c| } 
\hline
$\times \left( 1-\sin^2{\theta}\sin^2{\psi} \right)$ & Lepton Couplings\\
\hline
 $Z_{H_1^+}$ & $-\sin{\theta}/2 \left( 2\cos{\psi}\csc{\theta}\sin{\beta_1} - 2\cos{\theta}\sin{\beta_1}\sin^2{\psi}+\cos{\beta_1}\left( 2\cot{\theta}\sin{\psi}+\sin{2\psi} \right) \right)$ \\
\hline
 $Z_{H_2^+}$ & $\sin{\theta} \left(\left(\cos{\psi}+\cot{\theta} \right)\sin{\beta_1}\sin{\psi} + \cos{\beta_1}\left(-\cos{\psi}\csc{\theta} + \cos{\theta}\sin^2{\psi}  \right)  \right)$ \\ 
 \hline
$\sqrt{2}Z_{h_1}$ & $(1+k_{ff})\cos{\theta}\cos{\theta_3}\cos{\psi_3}+(1-k_{ff})\cos{\psi}\sin{\theta}\sin{\theta_3}$\\
\hline
$\sqrt{2}Z_{h_2}$ & $(k_{ff}-1)\cos{\theta_3}\cos{\psi}\sin{\beta_3}\sin{\theta}-(1+k_ff)\cos{\theta}\left(\cos{\psi_3}\sin{\beta_3}\sin{\theta_3}+\cos{\beta_3}\sin{\psi_3}  \right)$ \\
\hline
$\sqrt{2}Z_{h_3}$ & $(k_{ff}-1)\cos{\beta_3}\cos{\theta_3}\cos{\psi}\sin{\theta}+(k_{ff}+1)\cos{\theta}\left(\sin{\beta_3}\sin{\psi_3}-\cos{\beta_3}\cos{\psi_3}\sin{\theta_3}  \right)$ \\
\hline
$\sqrt{2}Z_{A_1}$ & $-\left(\left(1+k_{ff}\cos{2\theta} \right)\cos{\psi}\sin{\beta_2}+(1+k_{ff})\cos{\beta_2}\cos{\theta}\sin{\psi}     \right)$ \\
\hline
$\sqrt{2}Z_{A_2}$ & $-\left(1+k_{ff}\cos{2\theta} \right)\cos{\beta_2}\cos{\psi}+(1+k_{ff})\cos{\theta}\sin{\beta_2}\sin{\psi}$ \\
\hline
\end{tabular}}
\end{center}
Table $4$: These are the couplings to leptons appearing in equation \eqref{diagonal}.

\section{  Branching ratios and constraints on the charged Higgs bosons}

For the analysis of this section it is useful to write down the couplings of the charged Higgs to both, quarks and leptons in one single equation, see (\ref{Lquark}) and (\ref{diag}),
\begin{align}
\mathcal{L}= \frac{g}{\sqrt{2}M_W} \sum_{H^+=H_1^+,H_2^+} \left\lbrace  \xi_{H^+}^u \bar{u}_R V_{ud}m_u d_L -\xi_{H^+}^u \bar{u}_L V_{ud}m_d d_R - Z_{H^+} \bar{\nu}_L m_l l_R  \right\rbrace H^+ + h.c. \label{Lfermions}
\end{align}
Note that the charged Higgs interactions depend only on three mixing angles.   With the two different charged Higgs masses, this gives a five-dimensional parameter-space, as opposed to the 2HDM in which the interactions depend on the single charged Higgs mass and $\tan\beta \equiv v_2/v_1$.

Since the model has a close similarity to the lepton specific model, one can look at charged Higgs interactions in that model for guidance.   Logan and MacLennan \cite{Logan:2009uf} studied the constraints on the charged Higgs in this model, looking at direct LEP-II searches and flavor universality in tau decays.  Su and Thomas \cite{Su:2009fz} studied constraints from $b\rightarrow s\gamma$, $B^0-\bar{B}^0$ mixing and several other processes.   Aoki, Kanemura, Tsumura and Yagyu \cite{Aoki:2009ha}, in a very comprehensive study, considered these as well as $B\rightarrow\tau\nu$.   These papers and others are discussed in the review article of Ref. \cite{Branco:2011iw}, where it is noted that the bounds from radiative B decays and $R_b$ are the same as in the type I model, and bounds from flavor universality in tau decays are not significant unless $\tan\beta > 65$.   The most intriguing result is the possibility that the $\tau\nu$ decay mode of the charged Higgs can dominate the decay, {\it even above the $t\bar{b}$ threshold}.    For this to occur, one must have $\tan\beta > 10$.   Such fairly large values of $\tan\beta$ may have problems with perturbative unitarity \cite{Arhrib:2009hc}, although there are allowed regions of parameter-space with larger values of $\tan\beta$.    We will see that in the 3HDM, a dominance of the $\tau\nu$ decay mode can occur even if the ratio of vacuum expectations values is not particularly large.

\subsection{ Charged Higgs decays}

As noted above, the charged Higgs sector depends on three mixing angles and two masses.   As a first step, these parameters must be constrained by the requirement that they do not contradict LHC results on production and decay of the SM-Higgs like 125 GeV boson.    Given the number of parameters, a full $\chi$-squared analysis is unnecessary, and we will simply require that the couplings of the light Higgs ($h_1$ in Tables 1 and 2) be within $20\%$ of their Standard Model values for the $WW$, $ZZ$ and $tt$ couplings and $30\%$ for the $bb$ coupling\cite{Djouadi:2013qya}.    The $\gamma\gamma$ coupling will not provide useful information due to unknown contributions of heavy charged Higgs bosons (with arbitrary couplings) in the loop.    All of our results below will only consider regions in parameter-space that satisfy those constraints.

 We consider the branching ratios of the charged Higgs bosons. In this article we are going to assume that all of the additional neutral scalars are too heavy for  the charged Higgs to decay into them. Then the most relevant decay modes to consider are $H^+ \longrightarrow h W^+$, $\bar{b}t$, $\bar{\tau}\nu$ and possibly into $\mu \nu$. Focus will be placed mainly on heavy charged scalars $M_{H^\pm_i} \gg m_t$ and therefore the decay mode into a quark-antiquark pair will be dominated by the $\bar{b}t$ decay mode.
 The leading order expressions for the partial widths are given by 
  \begin{equation}
 \Gamma(H^\pm_{i}\rightarrow h W^\pm) = \frac{\sqrt{2} G_F}{16\pi}g_{H_i hW}^2 M_{H_i^+}^3 \left[ 1 + \frac{(M_W^2-m_{h}^2)^2}{M_{H_i^+}^4}- 2\frac{(M_W^2 + m_{h}^2)}{M_{H_i^+}^2} \right]^{3/2}, \label{hdecay} 
 \end{equation}
 \begin{equation}
 \Gamma (H_i^+ \rightarrow t \bar{b}) = \frac{3 G_F(\xi^u_{H_i^+})^2}{4 \pi \sqrt{2}} M_{H_i^+}  m_t^2 \left(1-\frac{m_t^2}{M_{H^+}^2}  \right)^2 , \label{quarkdecay}
 \end{equation}
 \begin{equation}
 \Gamma(H_i^+ \rightarrow \nu_l \bar{l}) = \frac{G_F}{4 \pi \sqrt{2}} M_{H_i^+}  \times \begin{cases}
      m_\tau^2 Z_{H_i^+}^2, & \ \tau \nu_\tau \\
      m_\mu m_\tau C_{H_i^+}^2, & \tau \nu_\mu , \ \mu \nu_\tau
    \end{cases} \label{leptondecay}
 \end{equation}
where $h=h_1$ being the SM-like Higgs boson is implied in \eqref{hdecay}. The $b$ quark and $\tau$ lepton masses have been neglected in \eqref{quarkdecay} and \eqref{leptondecay} respectively. For the leptonic decay \eqref{leptondecay} we included the flavor changing processes induced by \eqref{LFCNC} and the respective couplings, $C_{H_i^+}$, are given in the last two rows of table $3$.

In the lepton-specific 2HDM the couplings of the charged Higgs boson to quarks are proportional to the SM couplings multiplied by  $\cot{\beta}$. Also in that model, the couplings of the charged Higgs to right handed leptons are proportional to $\tan{\beta}$. Thus in the limit $\tan{\beta}\gg 1$, the charged scalar becomes quark-phobic but leptophilic enhancing the possibility of a large decay width into $\bar{\nu} \tau$, even above the $\bar{b}t$ threshold \cite{Branco:2011iw}.

Here we explore that possibility in the 3HDM. The couplings to the quarks in the 3HDM are given by 
 \begin{equation}
  \xi_{H_1^+}^u= \frac{\cos\beta_1 \cot\psi - \cos\theta \sin\beta_1}{\sin\theta}, \label{quarkcoupling1}
\end{equation}  
  \begin{equation}
\xi_{H_2^+}^u = -\frac{\cos\beta_1\cos\theta + \cot\psi\sin\beta_1}{\sin\theta}.\label{quarkcoupling2}
\end{equation}
We investigate quark-phobic points in the parameter space of the mixing angles $(\theta, \psi, \beta_1)$, i.e. points for which either $\xi_{H_1^+}^u$, $\xi_{H_2^+}^u$ or both are very small. Without loss of generality we do this analysis in the region of parameter space $ 0<( \theta, \psi, \beta_1) < \pi/2 $. The reason is that we want the vevs given in \eqref{angles} to be real and positive, so that the rotation angles $\theta, \psi$ can be restricted to be in the first quadrant. The reason to consider $\beta_1$ in that region is by noticing that taking $\beta_1 \rightarrow \beta_1 + \pi/2$ in \eqref{quarkcoupling1}  yields \eqref{quarkcoupling2}.

 By considering \eqref{quarkcoupling1} as a function of $\theta$ and $\psi$ and varying $\beta_1$ in that region, we find that there is always a surface for which  \eqref{quarkcoupling1} is equal to zero and that \eqref{quarkcoupling2} never crosses zero in that region. Thus the 3HDM allows for either $H_1$ or $H_2$ to be quark-phobic but not both at the same time. 

 When we consider the decay mode to the SM Higgs and a gauge boson, we see that the decay width \eqref{hdecay} is proportional to the square of the couplings $g_{H_i h W}$ given in table $1$. These couplings are dependent on the mixing angles $\theta_3$ and $\psi_3$, Therefore we also investigate points in the parameter-space that are both quark- and  gauge-phobic. We shall consider values of the mixing angles $\theta_3, \ \psi_3$ in the same region described above, although they could in principle be larger since they depend on the parameters of the scalar potential. By following the same procedure described above we find that it is always possible for \eqref{quarkcoupling1} and both gauge couplings  to cross zero at the same point. Therefore the 3HDM can always have a quark- and gauge-phobic $H_1$ decaying mostly into leptons and a gauge-phobic $H_2$ decaying most of the time to $tb$.

If one chooses parameters such that the ratio of vevs is large, one must consider unitarity and perturbativity. As remarked in the 2HDM review of Ref. \cite{Branco:2011iw}, having high ratios of vevs is only allowed for small regions of parameter space. Therefore  we consider points for which the ratios in equation \eqref{angles} are not too big ($\tan{\theta} ,\tan{\psi}< 6 $).

  The production cross section for charged Higgs bosons is dominated by the gluon-gluon fusion process $gg \rightarrow \bar{t}bH^+ $ and is proportional to the square of the couplings \eqref{quarkcoupling1} and \eqref{quarkcoupling2}. This mechanism will not produce charged Higgs bosons in the quark-phobic limit. In that case, they can still be produced by vector boson fusion (VBF) which is independent of quark couplings. As shown in Ref. \cite{Logan:2009uf}, the VBF production cross section is $2-3$ orders of magnitude smaller than the gluon-gluon fusion  cross section for $\tan{\beta}=1$, see figure $14$ of Ref \cite{Djouadi:2016eyy},  we thus consider quark-couplings which are $0, 0.03, 0.12$ times the $\tan{\beta}=1$ coupling.

A novel feature of this model is the appearance of the flavor changing decay modes given in \eqref{leptondecay}.   Usually the muonic decay of the charged Higgs is negligible due to the small muon coupling, however, here the tau coupling enters if the $\nu$ is a $\nu_\tau$. The region of parameter space for which the flavor changing branching ratios given in \eqref{leptondecay} are bigger than $10 \%$ is very small and is concentrated in the upper right corner of the plane $(\theta, \psi)$ which corresponds to large ratios of vevs. In the limit $\theta, \psi \rightarrow \pi/2$ 
these modes will be dominant with $H_1$ decaying $50\%$ into $\mu \nu_\tau$ and $\tau \nu_\mu$ each.

We find the branching ratios in figure $1$. We have chosen three benchmark points which give values for $\xi$ of $0, 0.03, 0.12$.   These points are $(\theta, \psi, \beta_1, \theta_3, \psi_3) $=$(1, 1.37, 0.36, 0.92, 1.14)$, $(1.30, 1.36, 0.59, 0.39, 1.36)$, $(1.41,1.41,0.25,0.28,1.41)$ respectively.    Changing the benchmark points, without changing the value of $\xi$ will not substantially alter these results.

For a quark Yukawa coupling of $0.12$ times the $\tan{\beta}=1$ 2HDM coupling, one sees that the branching ratio into $\tau \nu_{\tau}$ is substantial, although not dominant. As the quark Yukawa coupling gets smaller, the branching ratio into $\tau \nu_\tau$ becomes dominant for small masses.

 \begin{figure}[h]
 \centering
\includegraphics[scale=0.6]{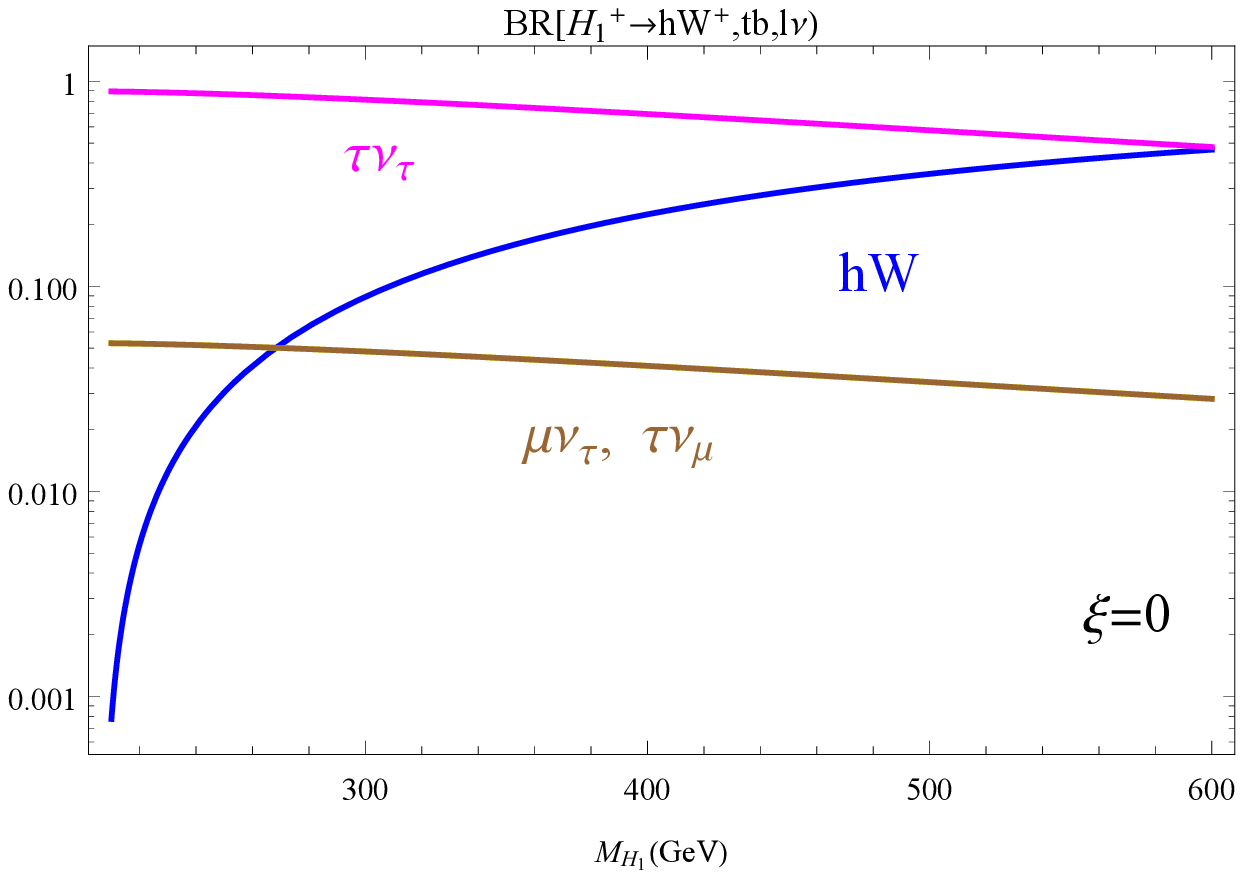}
\includegraphics[scale=0.6]{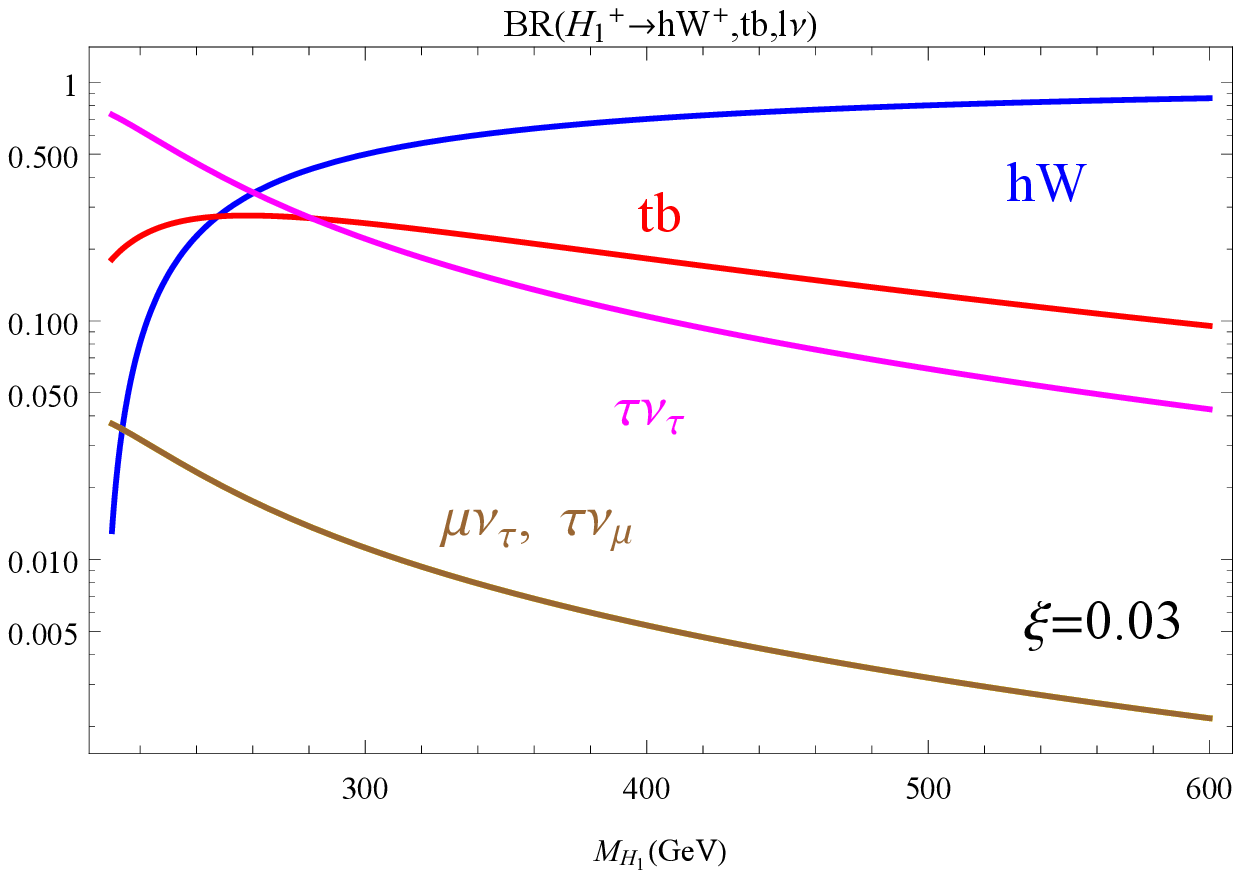}
\includegraphics[scale=0.6]{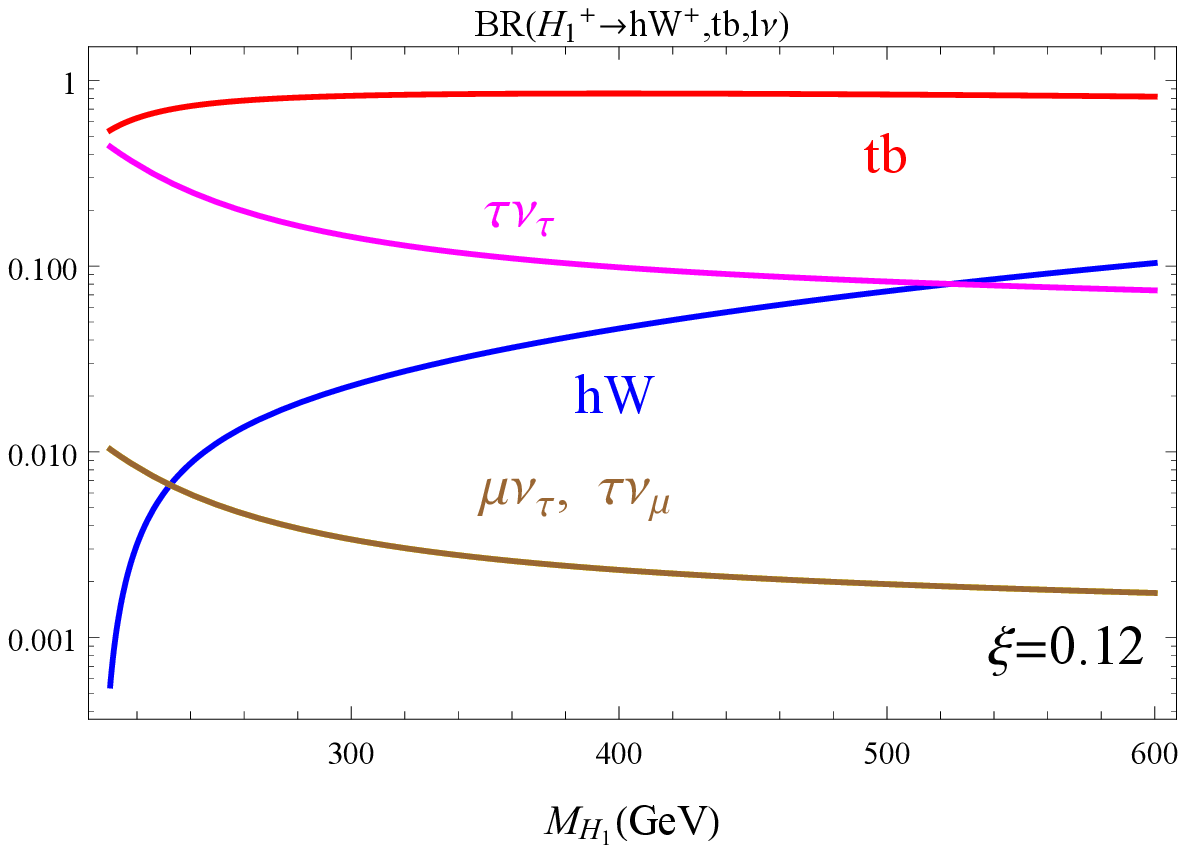}
\caption{Branching ratios of $H_1$ at the benchmark points in the text.   $\xi$ is the relative quark coupling to the charged Higgs from Eqs. 46 and 47.    $\xi=0$ corresponds to the pure quark-phobic limit.}
\end{figure}
 
The novel flavor changing mode discussed above, into $\mu \nu_\tau$ is small, but would be easier to detect. With the LHC operating with an integrated luminosity of $4000$ fb$^{-1}$ and  the pair production cross section by vector boson fusion (VBF), which is model independent, is around a few fb as can be seen from figure $8$ of Ref. \cite{Logan:2009uf}, producing approximately 300 events.    Note that a discussion of the neutral heavy Higgs decay into $\mu\tau$ can be found in Ref. \cite{Sher:2016rhh}.

	\subsection{ Constraints on the charged Higgs from B physics}

\subsubsection{ $B\rightarrow X_s \gamma$}
In this section we investigate bounds on the charged Higgs masses coming from the inclusive radiative decay $B\rightarrow X_s \gamma$ and give the leading order (LO) results at the matching scale $M_W$ at which the full theory is matched into an effective theory with five quark flavours in order to provide insight  on the effect of new physics, as is done in Ref. \cite{Carone:2009nu}. This is a well suited process to probe new physics. However it suffers from large theoretical uncertainties coming mainly from the choices of the renormalization and matching scales, which are not well defined at the leading order (LO) \cite{Ciuchini:1997xe}.
 
 The next to leading order calculations (NLO) in the SM  \cite{Chetyrkin:1996vx}, \cite{Greub:1996tg} and in the 2HDM \cite{Ciuchini:1997xe} are well known. The branching fraction of the inclusive radiative decay at NLO  is presented as \cite{Ciuchini:1997xe}
 \begin{align}
 \mathcal{B}(B \rightarrow X_s \gamma) = & \mathcal{B}(B \rightarrow X_c e \bar{\nu}_e)\Big|{\frac{V_{ts}^*V_{tb}}{V_{cb}}}\Big|^2 \frac{ 6 \alpha_{em}}{\pi f(z)\kappa(z)}\frac{\bar{m}_b^2(\mu_b)}{m_b^2}  \nonumber \\
 &\times (|D|^2+A)\left(1-\frac{\delta_{SL}^{NP}}{m_b^2}+\frac{\delta_\gamma^{NP}}{m_b^2}+\frac{\delta_c^{NP}}{m_c^2} \right),
 \end{align}
 where $z=m_c^2/m_b^2$ and $f$ and $\kappa$ are phase-space supression factors. The last term in parenthesis includes corrections obtained by the method of the heavy-quark effective theory (HQEFT) which relate the quark decay rate to the  hadronic process \cite{Neubert:1996qg}. The term $A$ is the correction coming from the Bremssthralung process $b\rightarrow s\gamma g$. Finally, $|D|^2$ contains the Wilson coefficents at the renormalization scale $\mu_b$ relevant to the radiative decay and is given by equation $(26)$ on Ref. \cite{Ciuchini:1997xe}. 
 As discussed in section $5$ of Ref. \cite{Ciuchini:1997xe} the LO 2HDM Wilson coefficients, to be added to SM ones, at the matching scale $m_W$ are given by 
\begin{equation}
\delta C_{7,8}^{(0)eff}(m_W) =  \frac{A_u^2}{3}F_{7,8}^{(1)}(y)- A_u A_d F_{7,8}^{(2)}(y),
\end{equation}
with
\begin{equation}
F_7^{(2)}(y)= \frac{y(3-5y)}{12(y-1)^2}+\frac{y(3y-2)}{6(y-1)^3}\ln{y},
\end{equation}
 \begin{equation}
 F_8^{(2)}(y)= \frac{y(3-y)}{4(y-1)^2}-\frac{y}{2(y-1)^3}\ln{y},
 \end{equation}
 \begin{equation}
 y=\frac{\bar{m}_t^2(m_W)}{M_H^2}
 \end{equation}
and $A_u$ and $A_d$ are the couplings of the charged Higgs to the quarks. In the type I 2HDM these are $A_u=A_d=\cot{\beta}$.

 To obtain the new Wilson coefficients of the 3HDM the only different thing is that we need to add the contribution from each $H_1^+$ and $H_2^+$ in the loop and modify the corresponding quark couplings. The result is 
 \begin{equation}
 \delta C_{7,8, NP}^{(0)eff}(m_W) =\sum_{i=1}^2 \left( \xi_{H_i^+}^u \right)^2 \left(\frac{1}{3}F_{7,8}^{(1)}(y_i)- F_{7,8}^{(2)}(y_i)\right), \quad y_i = \frac{\bar{m}_t^2(m_W)}{M_{H_i^2}} \label{LOcorrec}.
\end{equation}
Since the above equation is dependent on the three angles $(\theta, \psi, \beta_1)$ and the two charged Higgs masses we see that the branching fraction of the radiative decay will be dependent on five parameters.  The Wilson coefficients at the renormalization scale $\mu_b$ and at NLO cannot be summarized in a few lines and are extracted from Ref. \cite{Chetyrkin:1996vx}. 

Since we are interested in a region of parameter-space in which one of the charged Higgs bosons has suppressed couplings to quarks (we have checked that the suppression is sufficient to make the contribution negligible), the results will only depend on the mass of the other charged Higgs, $H_2^+$. For values of $\xi^u_{H^+_2} = 0.2, 0.6, 0.8$ and $1.4$, we find the results in Figure 2. These four points correspond to mixing angles $(\theta,\psi,\beta_1)$ = $(1.41,1.41,0.25)$, $(1.00,1.37,0.36)$, $(1.20,1.02,1.05)$ and $(0.69,1.21,0.46)$ respectively.       The first two of these correspond to the two values of the points listed in the previous subsection (corresponding to $\xi=0, 0.12$ (the other point in the last subsection gives results extremely similar to the $\xi=0.12$ point).    The other two points give results that are somewhat more significant for $B\to X_s\gamma$.

\begin{figure}[ht]
\centering
 \includegraphics[scale=0.9]{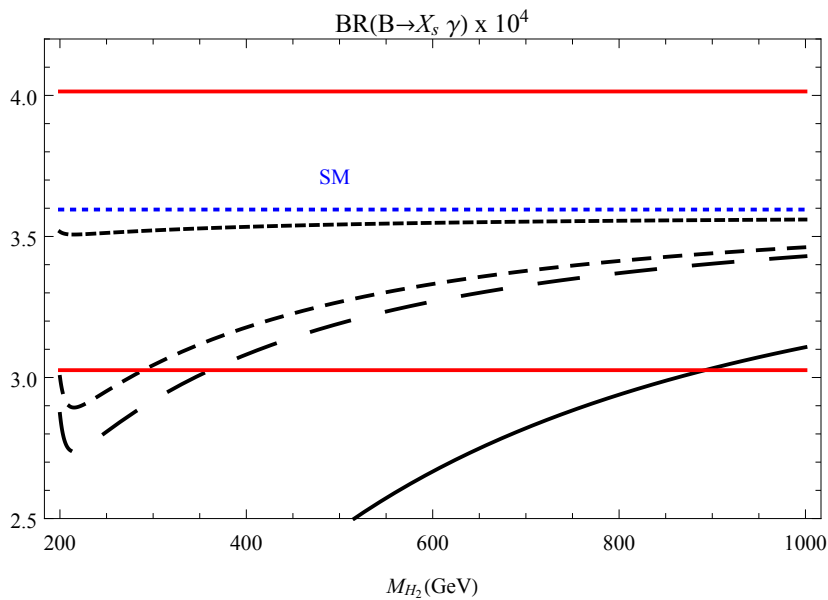}
\caption{Branching fraction for the radiative decay $B \rightarrow X_s \gamma$ as a function of the charged Higgs mass. The horizontal red lines give the $2\sigma$ experimentally allowed region and the blue dotted line give the central value of the SM prediction. The points of parameter space, in order of increasing dash length, correspond to values of $\xi^u_{H^+_2} = 0.2, 0.6, 0.8$ and $1.4$.}
\end{figure}

All black lines asymptote to the SM prediction $BR(B  \rightarrow X_s \gamma)=(3.6 \pm 0.36) \times 10^{-4}$ in the limit of high mass values.  The experimentally allowed region is $(3.52 \pm 0.23 \pm 0.09) \times 10^{-4}$ \cite{Barberio:2008fa}. The line corresponding to the first point lies entirely inside the $1 \sigma$ experimental region and therefore does not yield any bound.    Including NLO QCD corrections the lower bounds on the charged Higgs mass corresponding to the points considered in figure $2$ at $95 \%$ C.L. are given by $295$ GeV, $370$ GeV and $900$ GeV for the values of $\xi^u_{H^+_2} = 0.6, 0.8$ and $1.4$.    Thus, the bounds will be fairly weak unless $\xi^u_{H^+_2}$ is unusually large.     Note that a more recent analysis by Misiak et al. \cite{Misiak:2015xwa} does the NNLO calculation and finds results that are slightly lower, but well within a single standard deviation from the NLO results.

\subsubsection{ $R_b$}

In this subsection we focus on the observable
\begin{equation}
R_b = \frac{\Gamma(Z\rightarrow b \bar{b})}{\Gamma(Z \rightarrow hadrons)}
\end{equation}
which is sensitive to radiative corrections. It is shown in Ref. \cite{Haber:1999zh} that in non-minimal models containing only doublets, as is our case, the loop corrections due to virtual charged Higgs bosons always worsen agreement with experiment. In that same reference they introduce a parametrization for a general extended Higgs sector and calculate the contribution to $Zb\bar{b}$ from one-loop radiative corrections involving singly charged and neutral Higgs bosons. They obtained general expressions for the corrections to the left- and right-handed $Zb\bar{b}$ couplings, and then use the measurements of $R_b$ and $A_b$ (the coupling asymmetry) to constrain specific models.

 We write the interaction as 
 \begin{equation}
 \mathcal{L} \propto  
 Z_\mu \bar{b}\gamma^\mu \left[ \bar{g}_b^L P_L + \bar{g}_b^R P_R \right] b,
 \end{equation}
the effective couplings are then given by  $\bar{g}_b^{L,R}=g_{Zbb}^{L,R} + \delta g^{L,R}$, where $g_{Zbb}$ are the tree-level couplings and $\delta g$ are the radiative corrections.

The radiative corrections to SM extensions with only doublets and singlets, are given by the second term of equation $(4.5)$ of Ref. \cite{Haber:1999zh} 
\begin{equation}
\delta g^{L,R} = \pm \frac{1}{32 \pi^2}\frac{e}{s_W c_W} \sum_{i \neq G^+} \left( g_{H_i^+ }^{L,R} \right)^2 \times \left[ \frac{R_i}{(R_i-1)}- \frac{R_i \log{R_i}}{(R_i-1)^2} \right]
\end{equation}
where $s_W=\sin{\theta_W}$, $c_W=\cos{\theta_W}$ are the weak mixing factors, $R_i^2 = m_t^2/M_{H_i^+}^2$ and $g_{H_i^+}^{L,R}$ are obtained from the Lagrangian by writing the interaction of the charged Higgs to quarks as
\begin{equation}
\bar{t}\left( g^L P_L + g^R P_R \right) b H^+ + h.c.
\end{equation} 
The correction due to Goldstone boson exchange is excluded in the sum since is the same as in the SM. In the 3HDM the above coefficients are given by 
\begin{equation}
g_{H_i^+}^L = \sqrt{2} \frac{m_t}{v}\xi_{H_i^+}^u,
\end{equation}
\begin{equation}
g_{H_i^+}^R = - \sqrt{2} \frac{m_b}{v}\xi_{H_i^+}^u.
\end{equation}
Thus we can write the corrections as 
\begin{align}
\delta g^L = & \frac{1}{32 \pi^2}\frac{e}{s_W c_W}\left( \frac{\sqrt{2}m_t}{v} \right)^2 \sum_{i=1,2} (\xi_{H_i^+}^u)^2  \left[ \frac{R_i}{(R_i-1)}- \frac{R_1 \log{R_i}}{(R_i-1)^2} \right] 
\end{align}
and 
\begin{equation}
\delta g^R = -\frac{m_b^2}{m_t^2}\delta g^L.
\end{equation}

The experimentally allowed range is given by $R_b = 0.21642 \pm 0.00073$ \cite{Haber:1999zh}. We find bounds on the charged Higgs mass for the same quark-phobic points listed in figure $2$. However the first two points give a prediction that lies well inside the $2\sigma$ experimentally allowed region, therefore we only obtain a bound coming from the last point, namely
\begin{equation}
m_{H_2^+}> 395 \ \GeV  \ \ (95 \% \  C.L.).
\end{equation}
Thus we see that the radiative process $B \rightarrow X_s \gamma$ yields stronger constraints than those of $R_b$.

\subsubsection{Combining $B\rightarrow X_s\gamma$ and $R_b$}

In the above, we illustrated the bounds with a few benchmark points.   But there are only three angles and two masses, and we can scan the parameter-space to determine the allowed regions.   In the figures below we have plotted the regions in which the  $\chi^2$ tests for both $B \rightarrow X_s \gamma$ and $R_b$ processes are satisfied to $95 \%$ significance level. We parametrized the rotation angles  $\theta$ and $\psi$ in terms of the vevs $v_1, v_2$ and  imposed the constraint $\sum_i v_i^2=246 \GeV$.  In Figure 3, we show a contour plot of $v_1$ and $v_2$ for a specific value of $\beta_1$ chosen to give the quark-phobic point for $H_1$.

\begin{figure}[ht]
\centering
\includegraphics[scale=0.4]{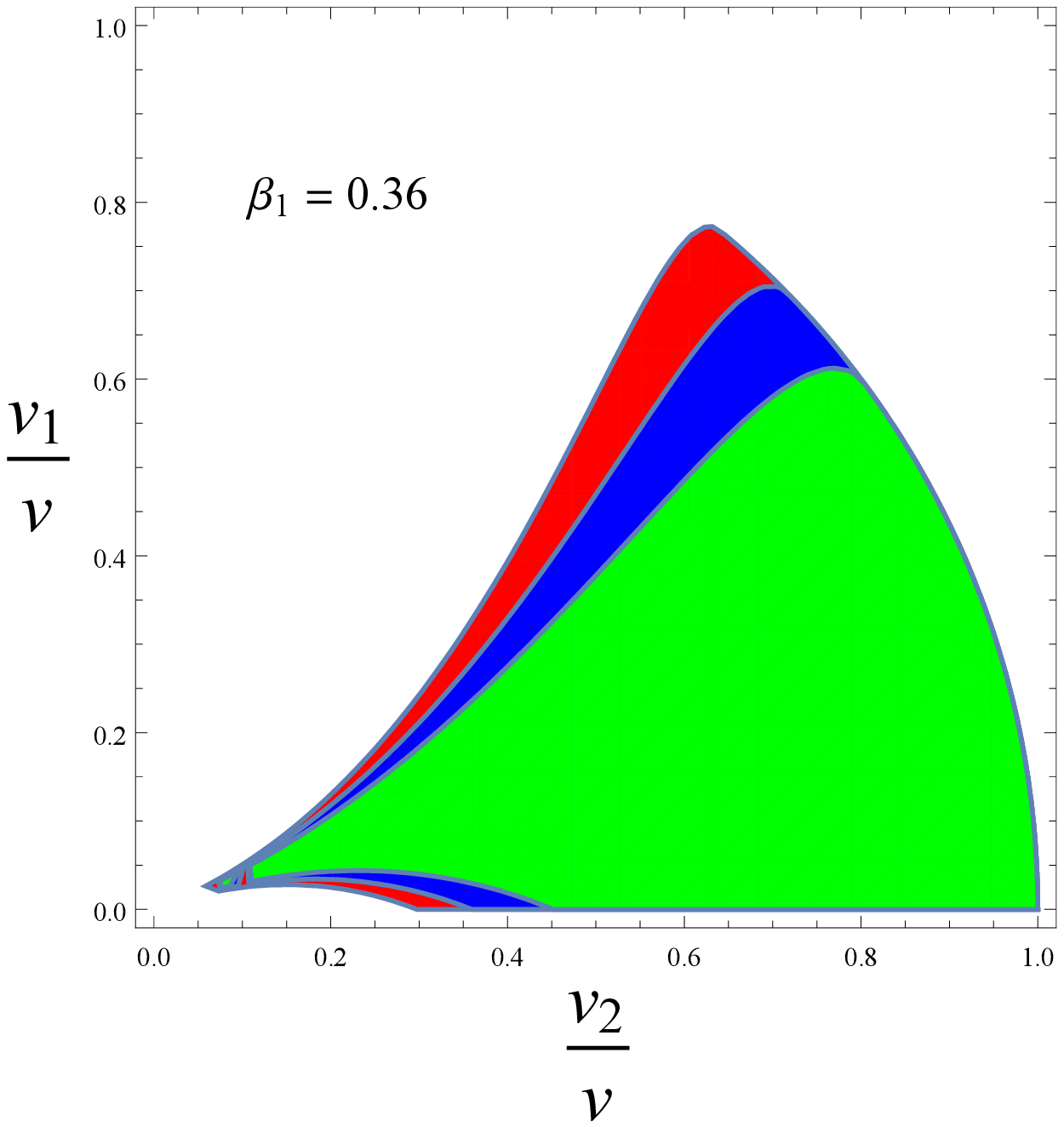},\includegraphics[scale=0.41]{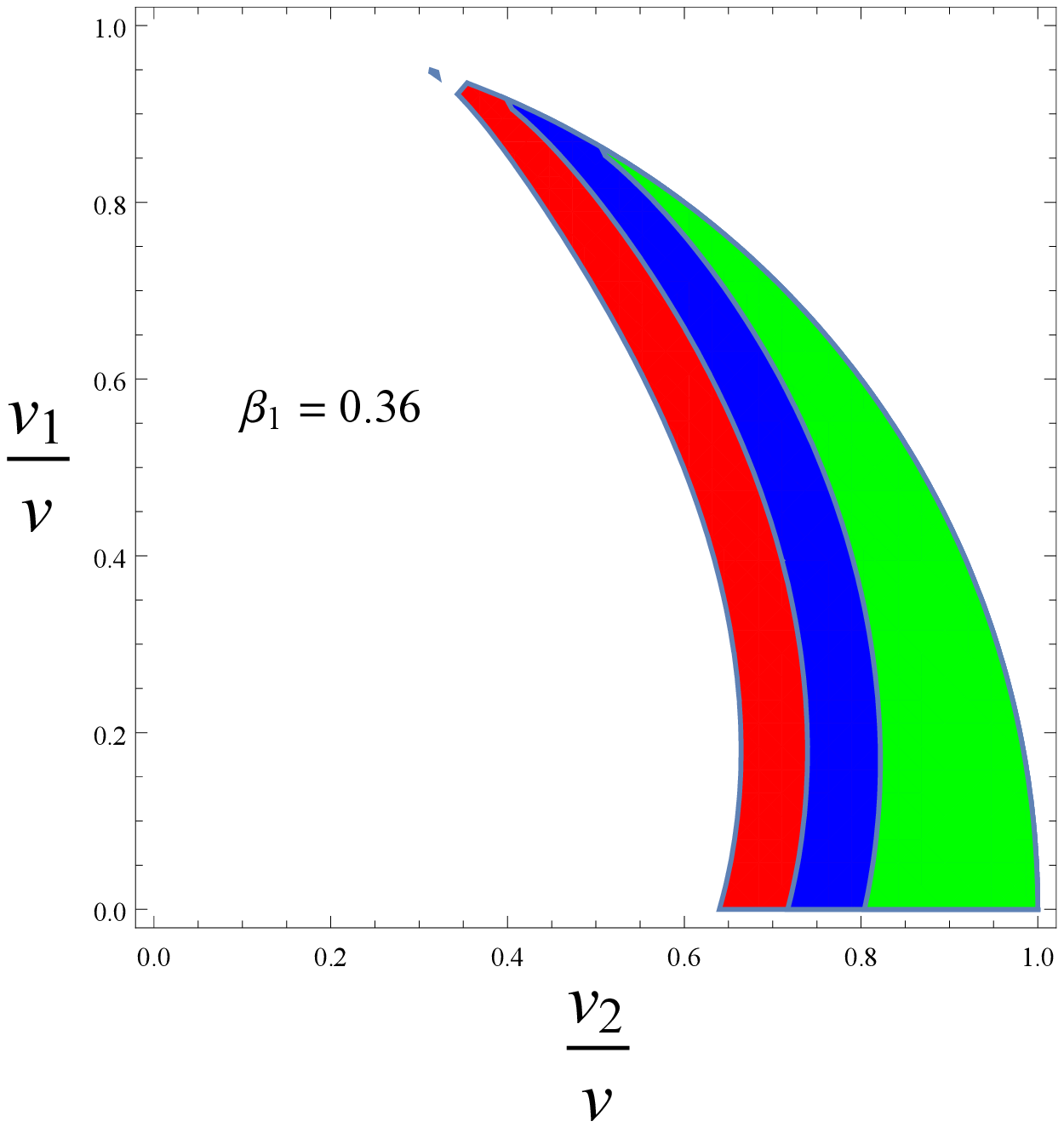}
\caption{Allowed region (colored) of the parameter space as a function of the vevs $v_1$ and $v_2$. The green, blue and red regions correspond to the mass values $M_H = 300, 500, 700 \ \GeV$ respectively.  We take the limit in which $H_2$ ($H_1$) effectively decouples from fermions on the left (right). The angle $\beta_1$ was chosen at the quark-phobic point of $H_1$.}
\end{figure}

We then consider different values of $\beta_1$ in Figure 4.   The patterns are clear.   Some regions of parameter-space, such as small $v_2$ and intermediate $v_1$, are excluded for all values of $\beta_1$.    This will become relevant when considering the possibility of $B\rightarrow \mu \nu_\tau$.   We now turn to the leptonic decays of the $B$

\begin{figure}[ht]
\centering
\includegraphics[scale=0.4]{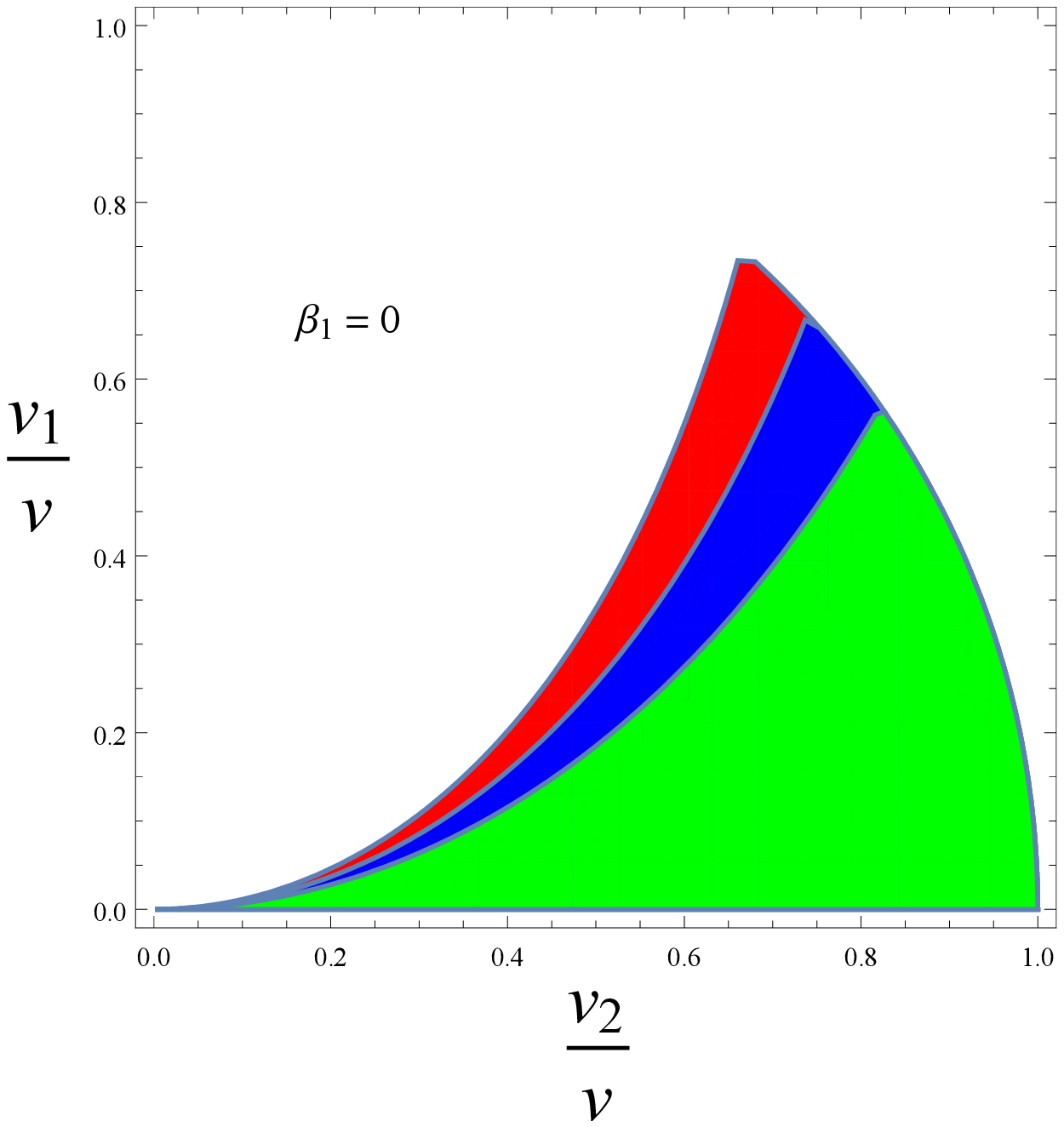},\includegraphics[scale=0.4]{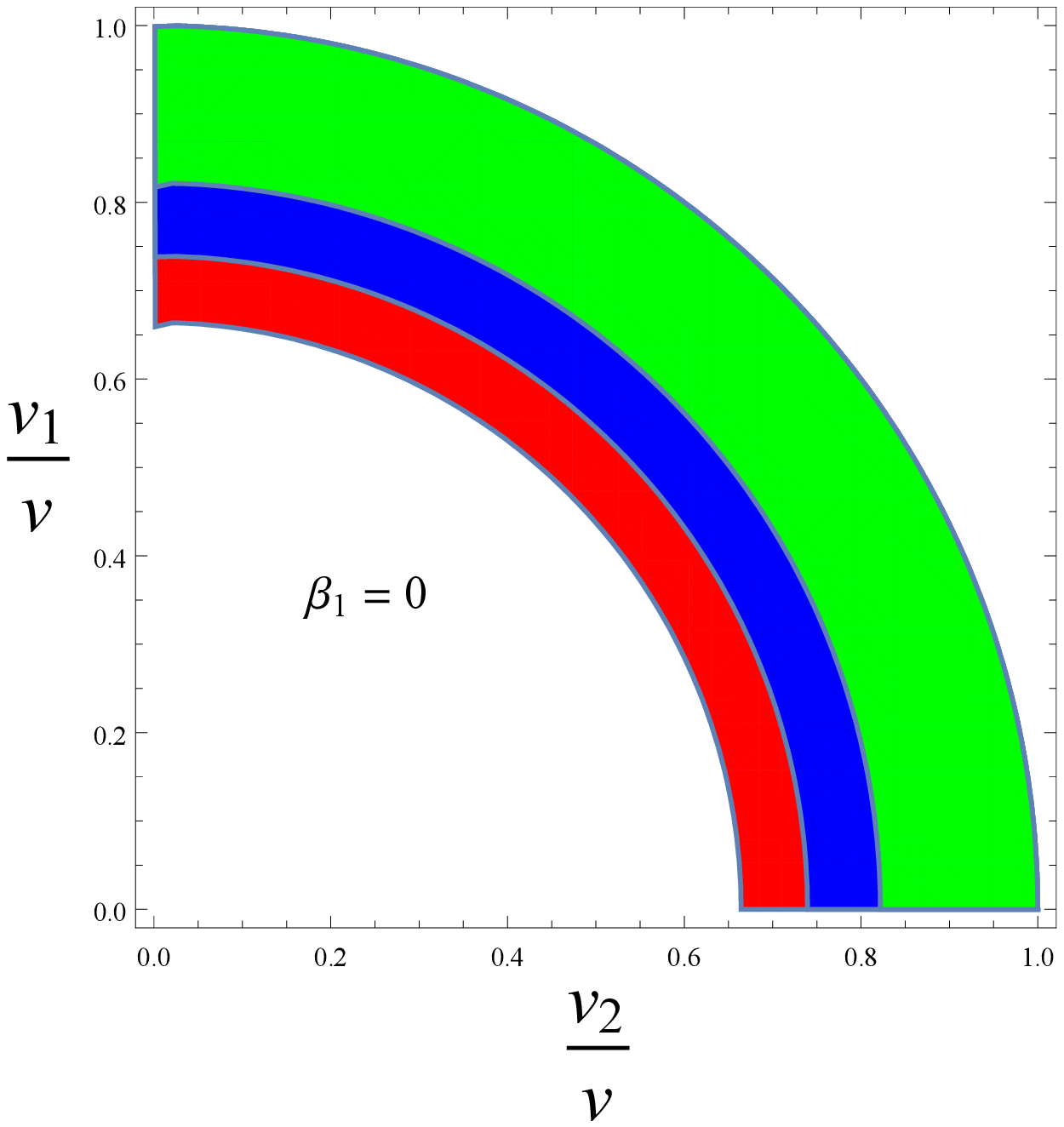}
\includegraphics[scale=0.4]{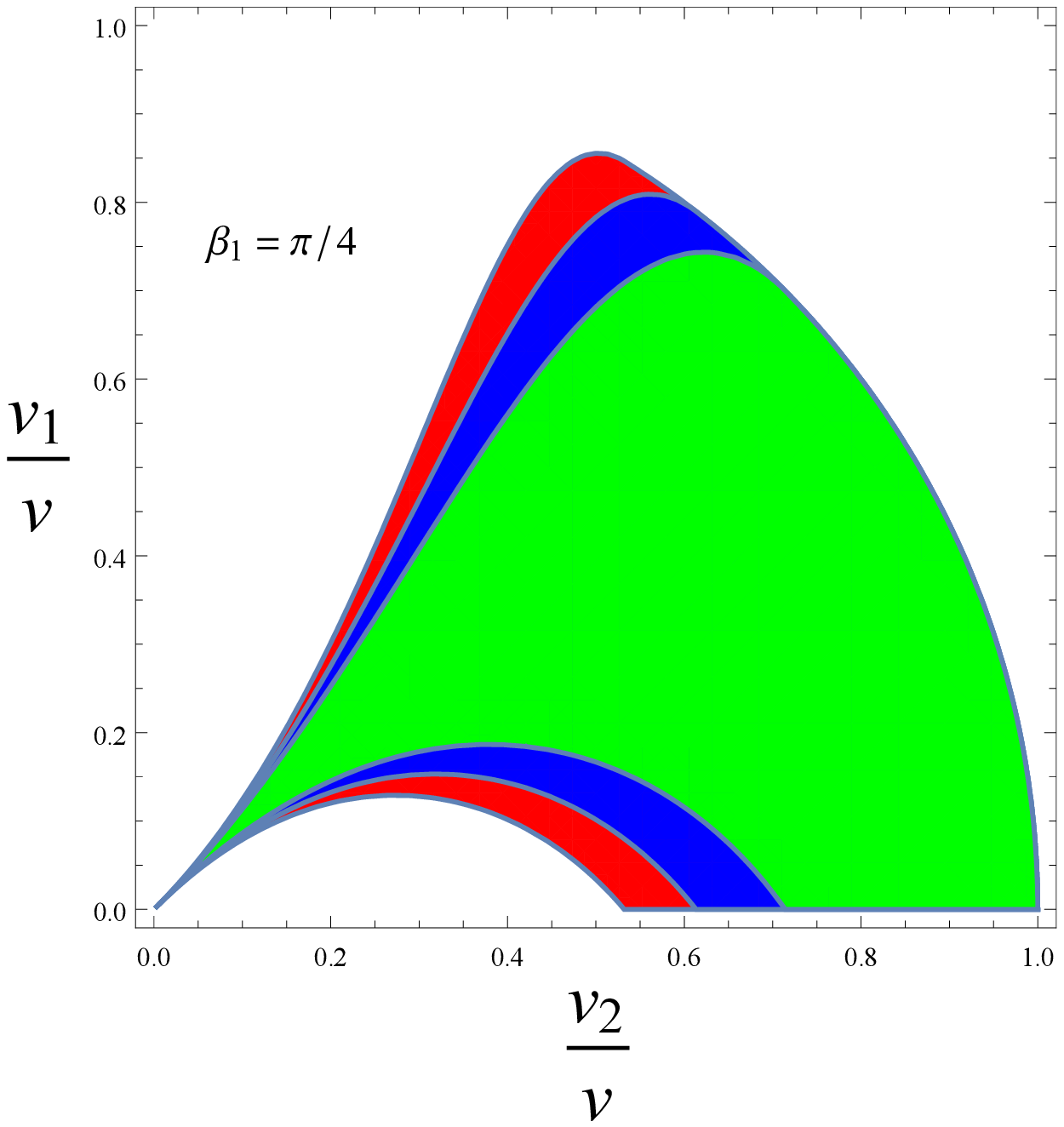},\includegraphics[scale=0.4]{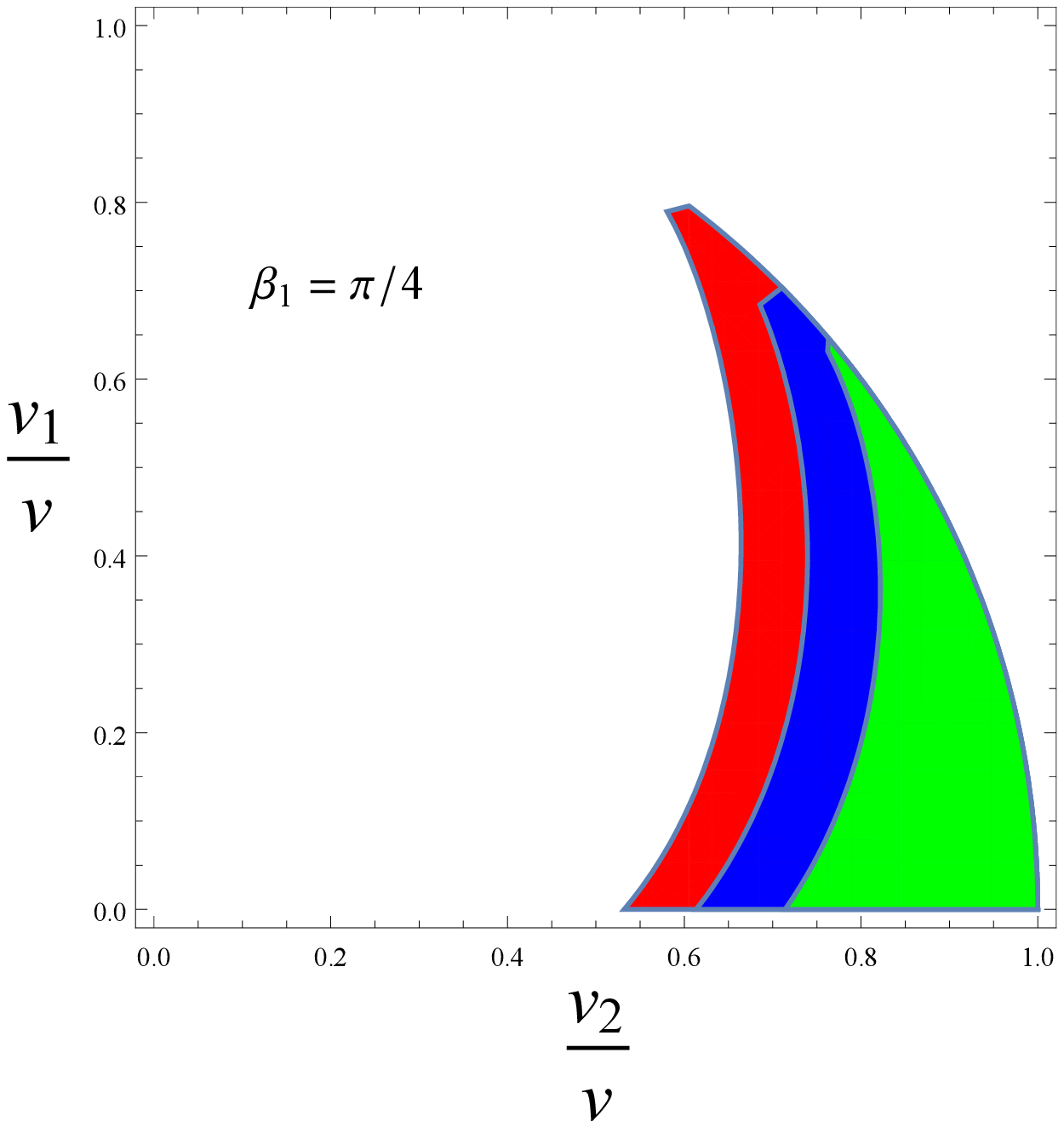}
\caption{Allowed region of parameter space for $\beta_1=0, \pi/4$. The charged Higgs $H_2$ ($H_1$) was taken very heavy on the left (right). The green, blue and red regions correspond to the mass values $M_H = 300, 500, 700 \ \GeV$ respectively. }	
\end{figure}

\subsubsection{ $B^- \rightarrow \tau \bar{\nu}_{\tau}$}
  
We study the charged current decay $B^- \rightarrow \tau \bar{\nu}$ type of modes which are just tree-level processess mediated by the electroweak gauge bosons $W^\pm$ and the charged Higgs bosons $H_i^\pm$.

Using the quark and lepton couplings of the charged Higgs mass eigenstates given by  \eqref{Lfermions}, the $W^\pm$ and $H_i^\pm$ effectively induce the four-Fermi interaction \cite{Hou:1992sy}
\begin{equation}
\mathcal{L} = -2\sqrt{2}G_{F}V_{u b} \left[ (\bar{u}\gamma^\mu P_L b )(\bar{l}\gamma_\mu P_L \nu_{l})   - R_{3HDM,l} (\bar{u}P_R b)(\bar{l}P_L \nu_{l})  \right],
\end{equation}
where
\begin{equation}
R_{3HDM,l} = \sum_{i=1}^2 \frac{m_{l}m_b}{M_{H_i^+}^2}Z_{H_i^+}\xi_{H_i^+}^u, \label{R1}
\end{equation}
where the first term give the SM contribution, while the second one give that of the charged scalars. In Ref. \cite{Grossman:1994jb} a study of multi-Higgs doublet models with natural flavor conservation was performed. In that article they assumed that all but the lightest of the charged scalars effectively decouple from fermions and carried out an analysis of phenomenological constraints on the Yukawa couplings. 

We do not make those assumptions here and instead modify the 2HDM result by the appropiate couplings and find that the charged Higgs bosons modify the SM expectation by the factor
\begin{equation}
 r_H \equiv \frac{BR(B^- \rightarrow l \bar{\nu})}{BR_{SM}(B^- \rightarrow l \bar{\nu})}  = \left\lvert 1 - \sum_{i=1}^2 m_B^2 \frac{\xi_{H_i^+}^u Z_{H_i^+}}{M_{H_i^+}^2} \right\rvert^2 
\end{equation}
which is independent of the lepton mass. We also call that factor $r_H$, following the notation in Ref. \cite{Hou:1992sy}.  Notice that in the limit $(\theta, \beta_1 ) \rightarrow  (\pi/2, 0)$, \eqref{quarkcoupling2} vanishes and $H_2$ becomes quark-phobic and the sum on the right-hand side collapses to 
\begin{equation}
\frac{BR(B \rightarrow l \nu)}{BR_{SM}(B \rightarrow l \nu)}  = \left\lvert 1 + \frac{m_B^2}{M_{H^+}^2} \right\lvert^2
\end{equation} 
which is the expression for the lepton-specific 2HDM. Therefore all the $B^- \rightarrow l \bar{\nu}$ modes are always enhanced in the lepton-specific 2HDM while they could be enhanced or supressed in the 3HDM by the same factor $r_H$. The ratio of the measured value to the SM prediction is $1.37 \pm 0.39$ \cite{Czarnecki:1998tn}. We compare this number with the $r_H$ prediction for the 3HDM and we find that in the quark-phobic points of  $H_1$ the 3HDM prediction lies completely inside the $1 \sigma$ region for all masses of $H_2$ starting from the threshold. Larger values  of the correction factor arise if we take the rotation angles $\theta, \psi \ll 1$, but that would correspond having a hierarchy on the vevs values $v_3 \gg v_1 \gg v_2 $. This is no surprise since it is  remarked in \cite{Hou:1992sy}, that for models where d-type quarks and charged leptons derive mass from different doublets there is no interesting effect. This is exactly  the case for lepton-specific-type models.

\subsubsection{ Flavor changing processess}
 
 This model has another interesting possibility that does not exist in any other versions of the 2HDM.    One can study $B\rightarrow \mu\nu_\tau$.    The Standard Model decay $B\rightarrow \mu\nu_\mu$ is very small due to helicity suppression, and the 2HDM charged Higgs contribution is negligible due to the small Yukawa coupling of the muon to the Higgs.   But in this model, the flavor-changing couplings are proportional to the geometric mean of the Yukawa couplings, and thus the $\tau$ Yukawa coupling can play an important role.   Of course, experimenters can not determine the flavor of the neutrino, so this would appear as a contribution to $B\rightarrow \mu\nu$.

 From \eqref{Hbasis} and \eqref{LFCNC} we can write down the flavor changing interaction of the charged Higgs bosons 
 \begin{equation}
 \mathcal{L}_{FCNC} \supseteq -\frac{\sqrt{2}}{v}\sqrt{m_\mu m_\tau} \sum_{H^+ = H_1^+, H_2^+} C_{H^+} \left[ \bar{\nu}_{\mu}P_R \tau + \bar{\nu}_\tau P_R \mu \right]H^+ +h.c.
 \end{equation}
where the $C_{H^+}$  are the flavor changing coupling constants given in the last two rows of table $3$. The charged Higgs induces the four-Fermi flavor-changing interaction 
\begin{equation}
\mathcal{L}_{4F} = \frac{4G_F}{\sqrt{2}} V_{ub} \sum_{H^+} R_{H^+}(\bar{u} P_R b) (\bar{\tau} P_L \nu_{\mu} + \bar{\mu} P_L \nu_{\tau}),
\end{equation}
where the sum is performed over the charged Higgs mass eigenstates and
\begin{equation}
R_{H^+} = \sqrt{m_\mu m_\tau}m_b \frac{\xi_{H^+}^u C_{H^+}}{M_{H^+}^2},
\end{equation}
similar to \eqref{R1}. The branching fraction for the flavor changing processes are given by the flavor conserving SM result times a correction factor, i.e.,
\begin{equation}
BR(B\rightarrow \mu \bar{\nu}_\tau,\tau \bar{\nu}_\mu) = BR(B\rightarrow \mu \bar{\nu}_\mu,\tau \bar{\nu}_\tau)_{SM} r_{H,l},
\end{equation}
where the $l$ index stands for the charged lepton and the correction factor is
\begin{equation}
r_{H,l}= \left(  \frac{\sqrt{m_\mu m_\tau}}{m_l}\sum_{H^+}\frac{m_B^2}{M_{H^+}^2}\xi_{H^+}^u C_{H^+} \right)^2.
\end{equation}
The SM branching ratio is given by \cite{Hou:1992sy}, \cite{Bona:2009cj}, \cite{Dingfelder:2016twb} 
\begin{equation}
BR_{SM}(B^- \rightarrow l^- \bar{\nu}) = \frac{G_F^2 m_B m_l^2}{8\pi} \left( 1-\frac{m_l^2}{m_B^2} \right)^2 f_B^2 |V_{ub}|^2 \tau_B.
\end{equation}

Using this formula the SM predictions are \cite{Bona:2009cj}
\begin{equation}
BR(B \rightarrow \tau \nu_\tau)_{SM} = (0.84 \pm 0.11)\times 10^{-4},
\end{equation}
\begin{equation}
BR(B \rightarrow \mu \nu_\mu)_{SM} = (3.8 \pm 0.5)\times 10^{-7}.
\end{equation}

 The Heavy Flavor Averaging Group (HFAG) found, as of July 2016,  the value $BR(B \rightarrow \tau \nu_\tau) = 1.06 \times 10^{-4}$  and the upper limit $BR(B \rightarrow \mu \nu_\mu)<1.0 \times 10^{-6} $ at $90 \%$ C.L. 

We investigate the region of the parameter space that allow for an enhancement of the flavor-changing decay  $B \rightarrow \mu \nu_\tau$ over the SM flavor-conserving prediction. One special case is the degenerate limit $M_{H^+_1}=M_{H^+_2}$, in which the correction factor vanishes trivially since the coupling constants satisfy the relation
\begin{equation}
\xi_{H_1^+}^u C_{H_1^+} =- \xi_{H_2^+}^u C_{H_2^+}
\end{equation}
for any rotation angle $(\theta, \psi, \beta_1)$. Thus one way to maximize the value of the correction factor is to take the decoupling limit in which either charged boson becomes extremely heavy and the other one take its threshold value. 

 The results are presented in figure $5$, where we consider the values of parameters in which the contribution from the charged Higgs is substantial.

\begin{center}
\includegraphics[scale=0.7]{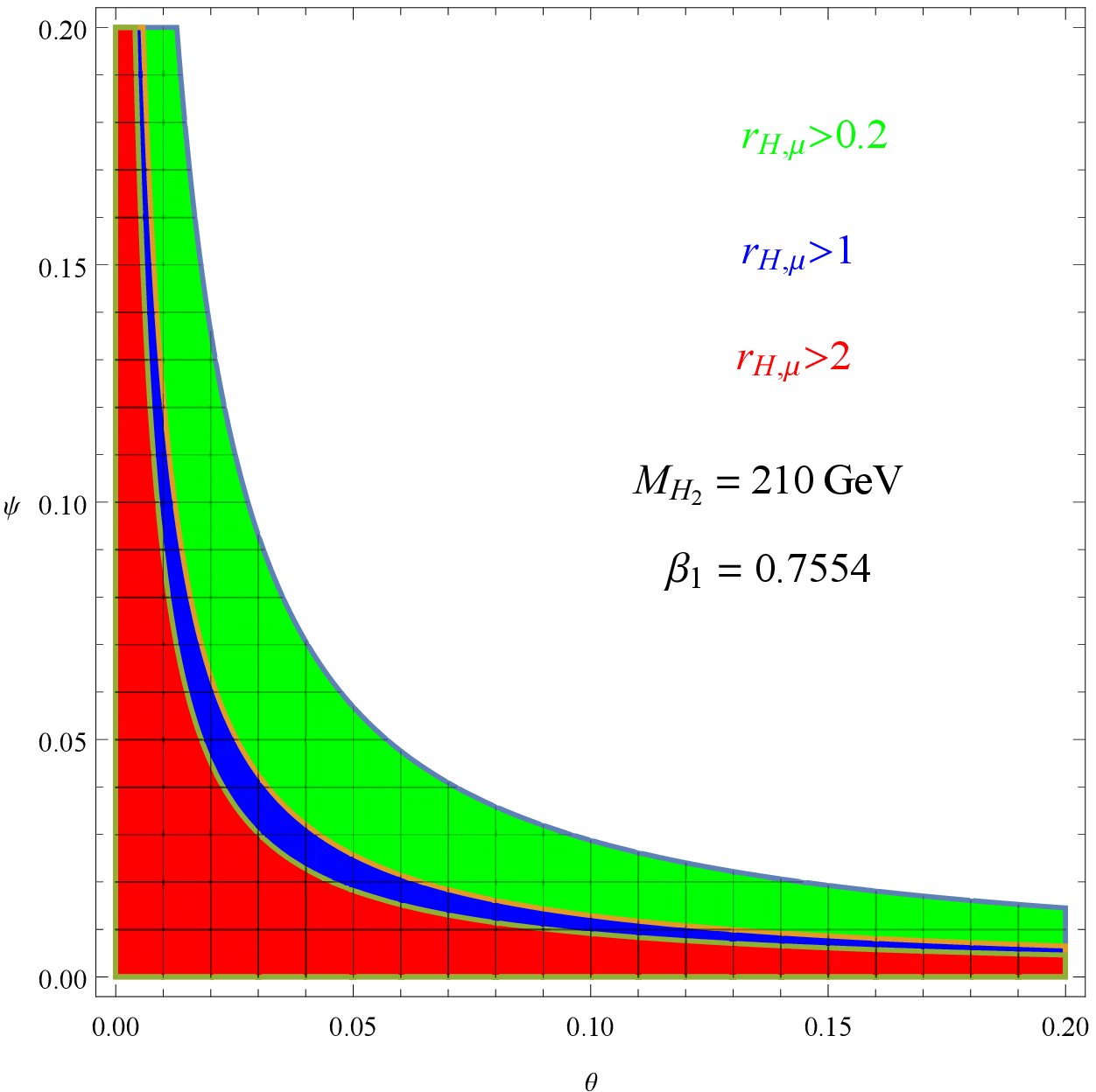}
\end{center}
Figure $5$: Region of the parameter space $(\theta, \psi)$ for which the correction factor $r_{H,\mu} >(0.2,1,2)$. \\

We have taken $M_{H_1^+}$ infinitely big and $M_{H_2^+}$ at its threshold value. The chosen value of $\beta_1$ maximizes the  area. \\

One natural question that arise is if  there exist a region of parameter space that allow a substantial enhancement and is permitted by the $\chi^2$ tests of the previous section. The answer is clearly negative as can be seen from the figure below where the colored region shows the  parameter space that is allowed by the radiative processes $B \rightarrow X_s \gamma$ and $R_b$ at $95 \%$ C.L. and the same values of $M_{H_2}$ and $\beta_1$ were used. That region is concentrated at the upper right corner and there is no overlapping between them.
\begin{center}
\includegraphics[scale=0.7]{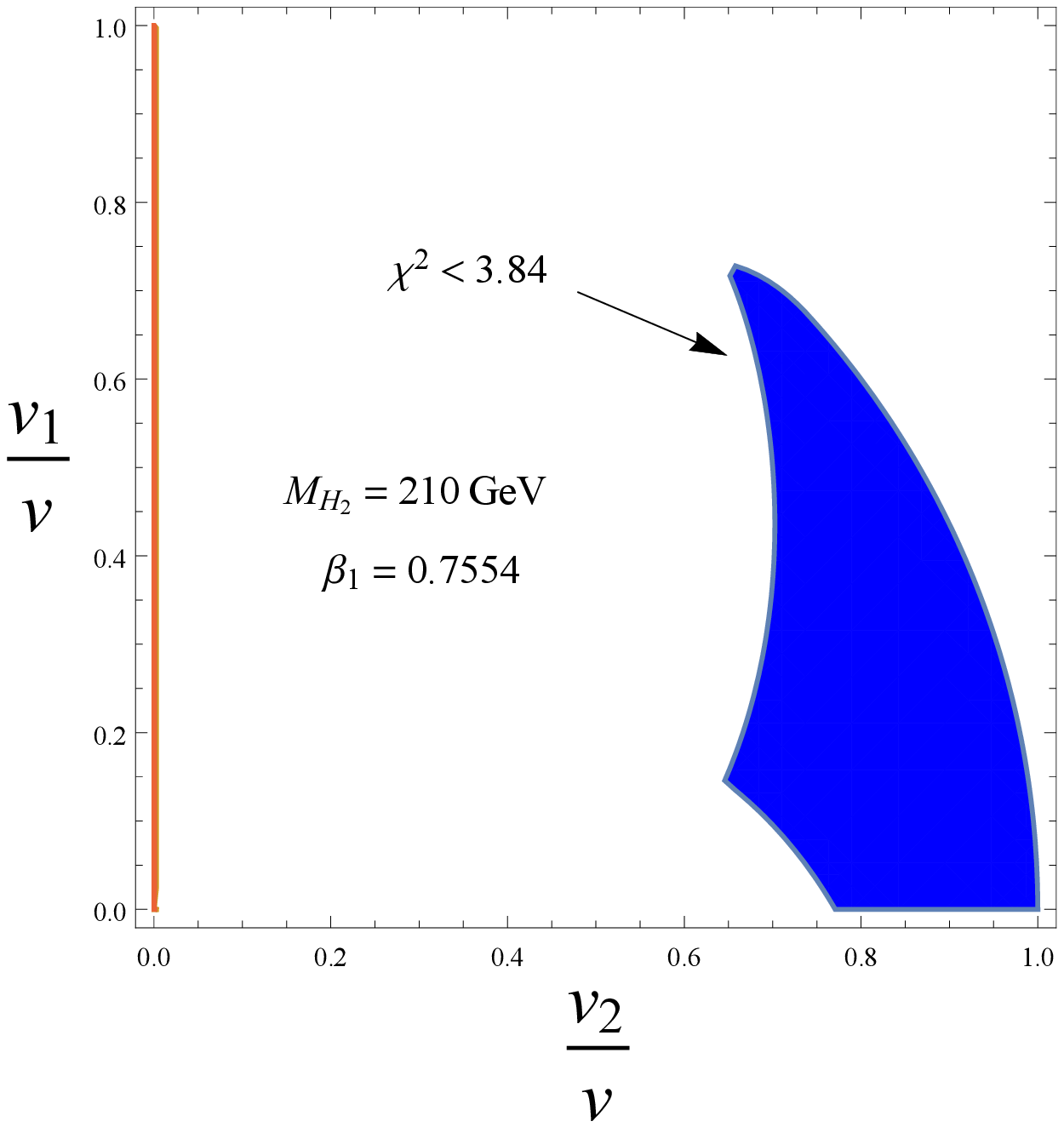},\includegraphics[scale=0.7]{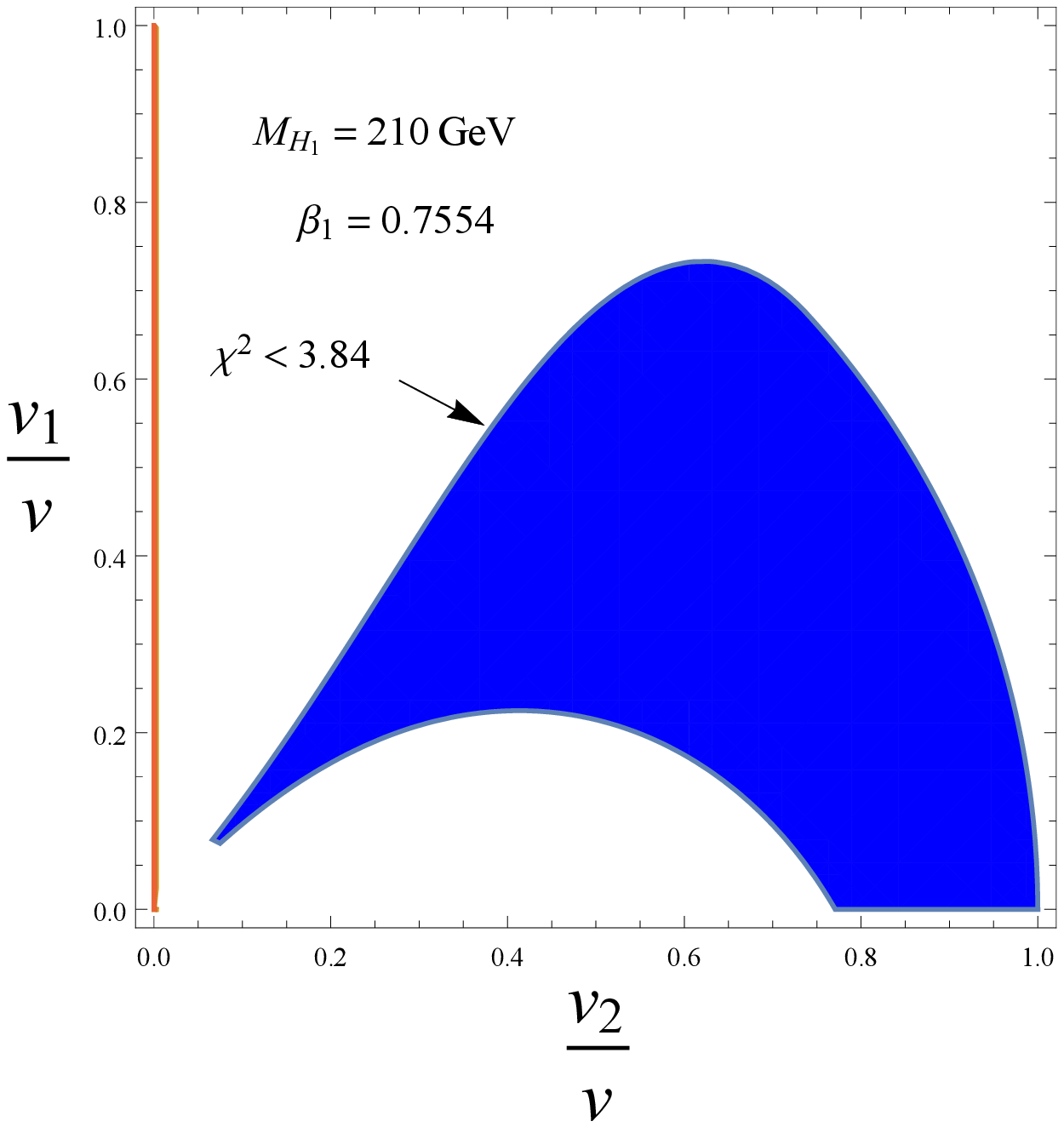}
\end{center}
Figure $6$: Allowed region of parameter space by the $\chi^2$ tests of the processes $B \rightarrow X_s \gamma$ and $R_b$. The decoupling limit of $H_1$($H_2$) has been taken in the left (right).

	\section{ Conclusions}

Motivated by hints of lepton flavor violation in Higgs decays and the strong constraints on tree level FCNC in the quark sector, we consider a 3HDM in which one doublet couples to quarks and the other two to leptons.   This structure can be imposed with a simple $Z_2$ symmetry.   The model has two charged Higgs pairs, two pseudoscalars and three scalars.   

The mass matrices and mixing angles are then determined.   In the charged Higgs sector, there are two masses and three mixing angles, two of which come from ratios of vevs.    We focus on the phenomenology of the charged Higgs bosons, concentrating on the region in which the lightest of the charged Higgs pairs is above the top quark mass.   The model is similar to the 2HDM lepton specific model.  In that model, the charged Higgs branching ratio into $\tau\nu$ can exceed that of $t\bar{b}$ only if the ratio of vevs ($\tan\beta$ in that model) is quite large, possibly large enough to raise unitarity concerns.    In this model, even without such a large ratio of vevs, the decays into $\tau\nu$, $t\bar{b}$ and $hW$ can all be comparable.    We also study the constraints from $B$ decay.   We find that there is a new decay, $B\rightarrow \mu\nu_\tau$, but the region of parameter-space in which this is substantial is inconsistent with other B decay bounds.

A unique feature of the model is the possibility of charged Higgs decay into $\mu\nu_\tau$.   The flavor of the neutrino can of course not be measured, but flavor does feed in through the size of the coupling, expected here to be the geometric mean of the muon and tau Yukawa couplings.    The decay branching fraction is typically a few percent, although for extreme regions of parameter-space can be much higher.    Although the branching fraction is small, the relative ease of detecting muons makes this decay worthy of careful study.

\section{Acknowledgments}

The authors would like to thank Chris Carone, Pedro Ferreira and Keith Thrasher for useful discussion.  This work was supported by the NSF under Grant PHY-1519644.

\newpage

\section*{Appendix A}

Minimizing the Higgs potential yields:
\begin{equation}
m_{11}^2= m_{13}^2 \frac{v_3}{v_1}-\lambda_{11}\frac{v_{1}^2}{2}-\frac{v_{2}^2}{2}(\lambda_{12}+\beta_{12}+\alpha_{12})-\frac{v_3^2}{2}(\lambda_{13}+\beta_{13}+\alpha_{13}),
\end{equation}
\begin{equation}
m_{33}^2= m_{13}^2 \frac{v_1}{v_3}-\lambda_{33}\frac{v_{3}^2}{2}-\frac{v_{2}^2}{2}(\lambda_{23}+\beta_{23}+\alpha_{23})-\frac{v_1^2}{2}(\lambda_{13}+\beta_{13}+\alpha_{13}),
\end{equation}
\begin{equation}
m_{22}^2=-\lambda_{22}\frac{v_{2}^2}{2}-\frac{v_{3}^2}{2}(\lambda_{23}+\beta_{23}+\alpha_{23})-\frac{v_1^2}{2}(\lambda_{12}+\beta_{12}+\alpha_{12}).
\end{equation}

The mass terms for the charged scalars are given by 
\begin{equation}
\mathcal{L}\supseteq(\phi_1^-,\phi_2^-,\phi_3^-)\phi_{matrix} \begin{pmatrix}
 \phi_1^+ \\
 \phi_2^+ \\
 \phi_3^+
                       \end{pmatrix},
\end{equation}
where the mass squared matrix for the charged scalars is given by 

\begin{equation}
\phi_{matrix} = \left( \begin{array}{ccc}
m_{13}^2 \frac{v_3}{v_1}-\frac{v_2^2}{2}A_{12}-\frac{v_3^2}{2}A_{13} & \frac{v_1 v_2}{2}A_{12} & -m_{13}^2+\frac{v_1 v_3}{2}A_{13} \\
\frac{v_1 v_2}{2}A_{12} & -\frac{v_1^2}{2}A_{12}-\frac{v_3^2}{2}A_{23} & \frac{v_2 v_3}{2}A_{23} \\
-m_{13}^2+\frac{v_1 v_3}{2}A_{13} & \frac{v_2 v_3}{2}A_{23} & m_{13}^2\frac{v_1}{v_3}-\frac{v_1^2}{2}A_{13}-\frac{v_2^2}{2}A_{23} \end{array} \right),
\end{equation}
and the following definitions where made
$$A_{12}=\alpha_{12}+\beta_{12},\quad
A_{13}=\alpha_{13}+\beta_{13},\quad
A_{23}=\alpha_{23}+\beta_{23}.$$
There is a zero eigenvalue corresponding to the charged Goldstone boson $G^{\pm}$ which gets eaten by the $W^{\pm}$. The mass squared of the charged Higgs particles is given by 
\begin{align} 
m_{\pm} = &\frac{1}{4 v_1 v_3}\left(2 m_{13}^2 v_{13}^2-v_1v_3 \left( v_1^2A_{13}+ v_2^2A_{23}+A_{12}v_{12}^2+(A_{13}+A_{23}) v_3^2\right) \right. \nonumber \\
    &\pm \left.\surd \left(-4 v_1 v_3 v^2 \left(A_{23} v_3^2 \left(-2 m_{13}^2+A_{13} v_1 v_3\right)+A_{12} v_1 \left(-2 m_{13}^2 v_1+A_{13} v_1^2 v_3+A_{23} v_2^2 v_3\right)\right)\right.\right. \nonumber \\
    & + \left.\left.\left(-2 m_{13}^2 \left(v_1^2+v_3^2\right)+v_1 v_3 \left(A_{13} v_1^2+A_{23} v_2^2+A_{12} \left(v_1^2+v_2^2\right)+(A_{13}+A_{23}) v_3^2\right)\right)^2\right)\right),
\end{align}
where we defined $v^2=v_1^2+v_2^2+v_3^2$,\quad $v_{ij}^2=v_i^2+v_j^2$, for $i,j=1,2,3$.
 
The mass terms for the pseudoscalars are given by 
\begin{equation}
\mathcal{L}\supseteq(\eta_1,\eta_2,\eta_3)\eta_{matrix} \begin{pmatrix}
 \eta_1 \\
 \eta_2 \\
 \eta_3
                       \end{pmatrix},
\end{equation}
with
\begin{equation}
\eta_{matrix} = \left( \begin{array}{ccc}
m_{13}^2 \frac{v_3}{v_1}-v_2^2 \beta_{12}-v_3^2\beta_{13} & v_1v_2\beta_{12}       & -m_{13}^2+v_1v_3\beta_{13}  \\
 v_1v_2\beta_{12}       &-v_1^2\beta_{12}-v_3^2\beta_{23} &v_2v_3\beta_{23}     \\
-m_{13}^2+v_1v_3\beta_{13} & v_2v_3\beta_{23} & m_{13}^2\frac{v_1}{v_3}-v_1^2\beta_{13}-v_2^2 \beta_{23} \end{array} \right),
\end{equation}
there is a zero eigenvalue corresponding to the neutral Goldstone boson $G^0$. The mass-squared of the physical pseudoscalar is 
\begin{align}
m_{A} = & -\frac{1}{2 v_1 v_3}\left(-m_{13}^2 v_{13}^2  +v_1 v_3 \left(v_1^2 (\beta_{12}+\beta_{13})+v_2^2 (\beta_{12}+\beta _{23})+v_3^2 (\beta_{13}+\beta_{23})\right) \right.  \nonumber\\ 
    &\pm \left.\surd \left(-4 v_1 v_3 v^2 \left(-m_{13}^2 \left(v_1^2 \beta_{12}+v_3^2 \beta _{23}\right)+v_1 v_3 \left(v_1^2 \beta_{12} \beta _{13}+v_2^2 \beta_{12} \beta_{23}+v_3^2 \beta _{13} \beta_{23}\right)\right) \right. \right.  \nonumber \\
    &+ \left.\left.\left(m_{13}^2 v_{13}^2-v_1 v_3 \left(v_1^2 (\beta_{12}+\beta_{13})+v_2^2 (\beta_{ 12}+\beta_{23})+v_3^2 (\beta_ {13}+\beta_{23})\right)\right)^2\right)\right).
\end{align}

Finally the mass terms for the scalars read
\begin{equation}
\mathcal{L}\supseteq(\rho_1,\rho_2,\rho_3)\rho_{matrix} \begin{pmatrix}
 \rho_1 \\
 \rho_2 \\
 \rho_3
                       \end{pmatrix},
\end{equation}
where the mass matrix is given by 
\begin{equation}
\rho_{matrix} = \left( \begin{array}{ccc}
m_{13}^2 \frac{v_3}{v_1}+v_1^2\lambda_{11} & B_{12}v_1v_2     & -m_{13}^2+B_{13}v_1v_3   \\
B_{12}v_1 v_2 &v_2^2 \lambda_{22} & B_{23}v_2v_3 \\
-m_{13}^2+B_{13}v_1v_3 & B_{23}v_2v_3 & m_{13}^2 \frac{v_1}{v_3}+v_3^2\lambda_{33} \end{array} \right),
\end{equation}
where we defined 
$$B_{12}=A_{12}+\lambda_{12}, \quad B_{13}=A_{13}+\lambda_{13} \quad B_{23}=A_{23}+\lambda_{23}.$$
There is no zero eigenvalue in this case and the scalar masses are given by the roots of the cubic characteristic equation of this matrix.

\section*{Appendix B: Vacuum stability or bounded from below conditions}

As the field value of each of the 12  components of the Higgs doublets go to infinity only the quartic terms of the potential given by  (\ref{potential}) become relevant. So we call this term 
\begin{align}
V_4 =& \frac{1}{2}\lambda_{11}(\Phi_1^\dagger \Phi_1)^2 +  \frac{1}{2}\lambda_{22}(\Phi_2^\dagger \Phi_2)^2+\frac{1}{2}\lambda_{33}(\Phi_3^\dagger \Phi_3)^2 \nonumber \\
   &+\lambda_{12}\Phi_1^\dagger \Phi_1\Phi_2^\dagger\Phi_2 + \lambda_{13}\Phi_1^\dagger \Phi_1\Phi_3^\dagger\Phi_3 + \lambda_{23}\Phi_2^\dagger \Phi_2\Phi_3^\dagger\Phi_3 \nonumber \\
   &+ \frac{\beta_{12}}{2}\left[(\Phi_1^\dagger \Phi_2)^2 + (\Phi_2^\dagger \Phi_1)^2\right] + \alpha_{12}\Phi_1^\dagger \Phi_2 \Phi_2^\dagger \Phi_1  \nonumber \\
   & + \frac{\beta_{13}}{2}\left[(\Phi_1^\dagger \Phi_3)^2 + (\Phi_3^\dagger \Phi_1)^2\right] + \alpha_{13}\Phi_1^\dagger \Phi_3 \Phi_3^\dagger \Phi_1  \nonumber \\
   &  + \frac{\beta_{23}}{2}\left[(\Phi_2^\dagger \Phi_3)^2 + (\Phi_3^\dagger \Phi_2)^2\right] + \alpha_{23}\Phi_2^\dagger \Phi_3 \Phi_3^\dagger \Phi_2. 
\end{align}
A simple way to obtain necessary conditions on the quartic parameters of the potential is to study its behaviour along specific field directions. 

Writting the three Higgs doublets as 
\begin{equation}
\Phi_1 = \left(\begin{array}{ccc}
 \phi_1 + i \phi_2\\
\phi_3 + i \phi_4
\end{array} \right), \quad \Phi_2 = \left(\begin{array}{ccc}
 \phi_5 + i \phi_6\\
\phi_7 + i \phi_8
\end{array} \right), \quad \Phi_3 = \left(\begin{array}{ccc}
 \phi_9 + i \phi_{10}\\
\phi_{11} + i \phi_{12}
\end{array} \right).
\end{equation}
We consider for example $\phi_3 \rightarrow \infty$ and $\phi_{11} \rightarrow \infty$ and we get 
\begin{equation}
V_4 = \phi_{11}^4 \left(
 \frac{\lambda_{33}}{2} + (\alpha_{13} + \beta_{13} + \lambda_{13})\left(\frac{\phi_3}{\phi_{11}}\right)^2 + \frac{\lambda_{11}}{2}\left(\frac{\phi_3}{\phi_{11}}\right)^4  \right).
\end{equation}
By making $\left(\frac{\phi_3}{\phi_{11}}\right)^2=x$, this can be seen as a simple polynomial of order two, which in order to be positive semi-definite, the following conditions 
\begin{equation}
\lambda_{33} \geq 0, \quad  \lambda_{11} \geq 0, \quad -\sqrt{\lambda_{11}\lambda_{33}} \leq \alpha_{13} + \beta_{13}+\lambda_{13}<0,
\end{equation}
should be satisfied. 
By doing similarly in all different  $\phi_i \phi_j$ planes we get the general necessary stability conditions
\begin{equation}
\lambda_{ii}\geq 0,
\end{equation}
\begin{equation}
-\sqrt{\lambda_{ii}\lambda_{jj}} \leq B_{ij}<0, \quad B_{ij}=\alpha_{ij} + \beta_{ij}+\lambda_{ij}
\end{equation}
\begin{equation}
-\sqrt{\lambda_{ii}\lambda_{jj}} \leq \bar{B}_{ij}<0, \quad  \bar{B}_{ij}=\alpha_{ij} - \beta_{ij}+\lambda_{ij}
\end{equation}

\end{document}